\documentclass[final, 12pt]{article}
\usepackage[utf8]{inputenc}
\usepackage{amsmath}
\usepackage{url}
\usepackage{graphicx}
\usepackage{amssymb}
\usepackage{amsmath,amsfonts,amssymb} 
\usepackage{graphicx} 
\usepackage{lmodern}
\usepackage{multicol}
\usepackage{hyperref} 
\usepackage{color}
\usepackage[mathscr]{euscript}
\usepackage{braket}
\usepackage[margin=1in]{geometry}
\usepackage{lscape}
\usepackage{caption}
\usepackage{subcaption}
\usepackage[toc]{appendix}
\usepackage{authblk}

\begin{document}
\title{\bf Closed trapping horizons without singularity}
\author{Pierre Binétruy${}^{1}$, Alexis Helou${}^{1,2}$, Frédéric Lamy${}^{1}$\footnote{frederic.lamy@apc.in2p3.fr} \vspace{0.8cm}} 

\affil{
${}^1$ \emph{AstroParticule et Cosmologie, Universit\'{e} Paris Diderot, CNRS, CEA,} \\
\emph{Observatoire de Paris, Sorbonne Paris Cit\'{e},} \\
Bât. Condorcet, 10 rue Alice Domon et L\'{e}onie Duquet, F-75205 Paris Cedex 13, France \\
${}^2$ \emph{Arnold Sommerfeld Center, Ludwig-Maximilians-Universität,} \\
Theresienstr. 37, 80333 M\"{u}nchen, Germany \vspace{0.2cm}} 
\date{April 10, 2018}
\maketitle

\renewcommand{\abstractname}{}
\begin{abstract}
\emph{We dedicate this article to the memory of Pierre Binétruy, who passed away on April 1, 2017. He was our guide in the journey through the darkest regions of the Universe.} 
\end{abstract}

\vspace{0.5cm} 

\renewcommand{\abstractname}{Abstract}
\begin{abstract}

In gravitational collapse leading to black hole formation, trapping horizons typically develop inside the contracting matter. Classically, an ingoing trapping horizon moves towards the centre where it reaches a curvature singularity, while an outgoing horizon moves towards the surface of the star where it becomes an isolated, null horizon. However, strong quantum effects at high curvature close to the centre could modify the classical picture substantially, e.g. by deflecting the ingoing horizon to larger radii, until it eventually reunites with the outgoing horizon. We here analyse some existing models of regular ``black holes'' of finite lifespan formed out of ingoing null shells collapsing from $\mathscr{I}^-$, after giving general conditions for the existence of (singularity-free) closed trapping horizons. We study the energy-momentum tensor of such models by solving Einstein's equations in reverse and give an explicit form of the metric to model a Hawking radiation reaching $\mathscr{I}^+$. A major flaw of the models aiming at describing the formation of black holes (with a Vaidya limit on $\mathscr{I}^-$) as well as their evaporation is finally exhibited: they necessarily violate the null energy condition up to $\mathscr{I}^-$, i.e. in a non-compact region of spacetime.

\end{abstract}

\section{Introduction}

The usual theoretical tool for describing a black hole geometry is the Schwarzschild metric, which is still widely used a hundred years after its discovery. It represents a static and eternal black hole. Already at the classical level, this static geometry does not address all the complexity of black hole formation. In collapse to form a black hole, a first (marginally) trapped surface forms inside the contracting matter, at the location where the expansion of outgoing null geodesic congruence $\theta_+$ changes sign, from positive to negative. Subsequently, this first marginally outer trapped surface (MOTS) develops into a 3D hypersurface, or marginally outer trapped tube (MOTT), in two directions: an outgoing component, which moves towards the surface of the collapsing cloud and becomes a null, isolated horizon once it reaches it, and an ingoing component, which moves towards the centre where the classical singularity forms \cite{Booth:2005ng,Helou:2016xyu}. This general dynamical picture is observed in a number of known analytic solutions to Einstein equations such as the ones of Oppenheimer-Snyder, Vaidya, and Lemaître-Tolman-Bondi, where the first MOTS forms at the surface, at the centre, or in the bulk, respectively.
 
 However, even these more refined dynamical geometries remain purely classical, and cannot be fully satisfying. Indeed the central singularity they display can be seen as an incompleteness of General Relativity and should be taken care of by a quantum theory of gravity. Moreover, the back-reaction of Hawking radiation should be taken into account in a fully dynamical description of black holes. It should be noted that avoiding the classical singularity requires violating at least one of the assumptions of the Penrose singularity theorem \cite{Penrose:1965}, namely global hyperbolicity, the weak energy condition or the existence of a trapped surface. The side we will be taking in this article, dealing with dynamical geometries, consists in violating the weakest of energy conditions, the null energy condition\footnote{See Eq.\eqref{NEC} for a definition of the NEC.}. \\
 How to further proceed? In the absence of a quantum theory of gravity, an option consists in building toy-models for the formation and evaporation of non-singular ``black holes''  with resolved singularity, through the use of effective metrics which include quantum effects. \\
 
One can think of some possible effective descriptions of these quantum effects, in preparation for the day we find a quantum theory of gravity, as pointed out in \cite{Casadio:2016fev}. We usually work with the outgoing horizon, and the semi-classical Hawking radiation that goes with it, but we most often completely overlook the role of the ingoing horizon (see however \cite{Booth:2005ng,Helou:2016xyu}), because it is thought to be hidden behind the outgoing horizon. But what if the shrinking dynamics of the evaporating outer horizon were to reveal the  inner regions of the black hole, and make the ingoing horizon an observable part of the collapse \cite{Hiscock:1980ze,Hiscock:1981xb}? And what if unknown quantum effects in the high curvature region close to the centre were sufficient to modify the trajectory of the ingoing horizon, preventing it from reaching the centre and making it bounce towards the outer horizon \cite{Bardeen:2017ypp}? This last scenario would give a new perspective on black hole evaporation, with an effect coming from inside the black hole and not from the emitting outer horizon. In the present paper, we study the shrinking of the outer horizon due to Hawking radiation, as well as the less common quantum bounce of the inner horizon in the high curvature region\footnote{Another motivation for studying these models is the implication of the absence of singularity on the so-called black hole information paradox. This will not be studied in details in the present paper.}.

We will here be working in spherical symmetry. In this context, the classical singularity lies at the centre $R=0$, where $R$ is the areal radius of the 2-spheres of symmetry. We also have a notion of gravitational energy, the Misner-Sharp mass $M(R)$, denoting the mass and gravitational energy enclosed in a sphere of radius $R$. Although there is no preferred foliation to define our trapped surfaces \cite{Wald:1991zz}, we here choose to use the symmetry of the spacetime \cite{Faraoni:2016xgy} and follow the standard way which consists in working with the spherically symmetric horizons (see also \cite{Helou:2016xyu} for motivations). We then call a ``quasi-local horizon'' the locus where $R=2M(R)$ (a MOTS, $\theta_+=0$), and a ``trapped region'' the connected part of spacetime where $R<2M(R)$ \cite{Dafermos:2004wr}\footnote{We use the convention $G=c=1$.}. 

When trying to regularize the black hole central singularity, a first idea is to modify the Schwarzschild metric while conserving its asymptotic structure and event horizon, in order to get a static and eternal, but singularity-free, black hole. A possibility is to ask for a de Sitter behaviour close to the centre \cite{Frolov:1988vj,Dymnikova:1992} interpreted as an effect of a Planckian cutoff \cite{Hayward:2005gi}, or to use non-linear electrodynamics as a source \cite{AyonBeato:1998ub} (see also \cite{Ansoldi:2008jw} and references therein).

Another possibility is to focus on the quasi-local horizon -- which is better-suited for dynamical situations such as gravitational collapse -- and see if one could use its dynamics to regularize the singularity. Hawking \cite{Hawking:2014tga} expressed the possibility that a true event horizon may never form, only ``apparent horizons which persist for a period of time'', and therefore that ``there are no black holes'', in the sense of causally disconnected regions of spacetime. This idea is not new, and has been pioneered by Frolov and Vilkovisky \cite{Frolov:1979,Frolov:1981mz}, Roman and Bergmann \cite{Roman:1983zza}, and Hajicek \cite{Hajicek:1986hj,Hajicek:1986hn}, in the eighties. More recently, Hayward applied this idea to his trapping horizons \cite{Hayward:2005ny,Hayward:2005gi}. He obtained a regular black hole\footnote{For an analysis of the static version of Hayward's regular black hole with the tools of the so-called Horizon Quantum Mechanics (HQM), see \cite{Giugno:2017xtl}. For a general introduction on the HQM, see \cite{Casadio:2016fev}.} with closed trapping horizons, \emph{i.e.} two trapping horizons forming from the first MOTS and merging together into a last MOTS, in an asymptotically Minkowski spacetime without any event horizon (see also \cite{Hossenfelder:2009fc} for a variation on this idea). Around the same time, Ashtekar and Bojowald also proposed a similar model \cite{Ashtekar:2005cj}. Even more recently, these ideas attracted new interests: Frolov presented other models with closed trapping horizons \cite{Frolov:2014jva, Frolov:2014wja, Frolov:2016pav}, as well as Bardeen \cite{Bardeen:2014uaa, Bardeen:2017ypp}. In this paper, we will give examples of explicit metrics for the Hayward, Frolov and Bardeen spacetimes and discuss these various models.

Rovelli \emph{et al.}~suggested to use the same bounce as in Loop Quantum Cosmology -- that resolves the Big Bang singularity -- but applied to black holes, and called the resulting regular objects ``Planck stars'' \cite{Rovelli:2014cta,DeLorenzo:2014pta}. This is a similar idea, but it adds the assumption that the matter should bounce at critical density, therefore turning the black hole region into a white hole one. Finally, a related, recent idea suggests that the underlying causality of the spacetime should become non-dynamical (Minkowski-like) at Planckian energy-scales, allowing for regular bounces between black hole and white hole behaviours of the stellar object \cite{Barcelo:2014cla,Carballo-Rubio:2015uwa}.

The present article is organized as follows. In Section \ref{section_Roman}, we study general conditions to obtain singularity-free spacetimes with closed trapping horizons, and give some examples taken from the literature. In Section \ref{section_globalprop}, we focus on the behaviour of null geodesics in these models, and define some relevant regions to investigate their phenomenology. Finally in Section \ref{section_energy}, we solve Einstein's equations in reverse to obtain the expression of the energy-momentum tensor for these models and analyse the weakest of energy conditions, the null energy condition (NEC). We find an explicit metric that recovers a null outgoing fluid mimicking Hawking radiation on $\mathscr{I}^+$, without having to make junctions. We ultimately show that all models based on the collapse of ingoing null shells, hence (asymptotically) described by a Vaidya metric on $\mathscr{I}^-$, and willing to describe Hawking's evaporation, are doomed to violate the energy conditions in a non-compact region of spacetime.

\section{Singularity-free spacetimes with closed trapping horizons}
 \label{section_Roman}
\subsection{Trapping horizons in classical analytic collapse}
A very general feature of collapses leading to black holes is the formation of trapping horizons \cite{Hayward:1993wb}. These are foliated by 2D marginally outer trapped surfaces (MOTS), which are also called apparent horizons\footnote{Concerning the nomenclature of quasi-local horizons, we will here follow \cite{Poisson,Visser:2014zqa} in not making the distinction between the 3D-trapping horizon of \cite{Hayward:1993wb}, where $R=2M$, and the 2D-apparent horizon \cite{Hawking:1973uf} which foliates it (see \cite{Helou:2016xyu} for details). As a result, the term trapping horizon will be mainly used in the following.}. In the course of the collapse, a first MOTS will appear, and trapped surfaces will then develop. In the usual, analytic black hole spacetimes, the location of this first\footnote{Using the time-slice of the comoving observer \cite{Helou:2016xyu}.} MOTS is known. It appears at the surface in the Oppenheimer-Snyder (OS) homogeneous dust collapse (see middle panel of Fig.~\ref{fig:Roman_Bergmann}), in the bulk of the collapsing matter for some classes of Lemaître-Tolman-Bondi (LTB) spacetimes, or at the centre in Vaidya null-dust collapse (left panel of Fig.~\ref{fig:Roman_Bergmann}) as well as in some other classes of LTB spacetimes \cite{Booth:2005ng,Helou:2016xyu}. When it is not formed at the centre, it immediately separates into an ingoing apparent horizon and an outgoing one, where ingoing/outgoing  refers to the motion with respect to the collapsing matter (this is a hydrodynamical concept, not to be confused with the geometrical one of inner/outer trapping horizons).
    
\subsection{Closed trapping horizons}
\label{section2.1}
\begin{figure}[h!]
\centering
  \includegraphics[scale=0.3]{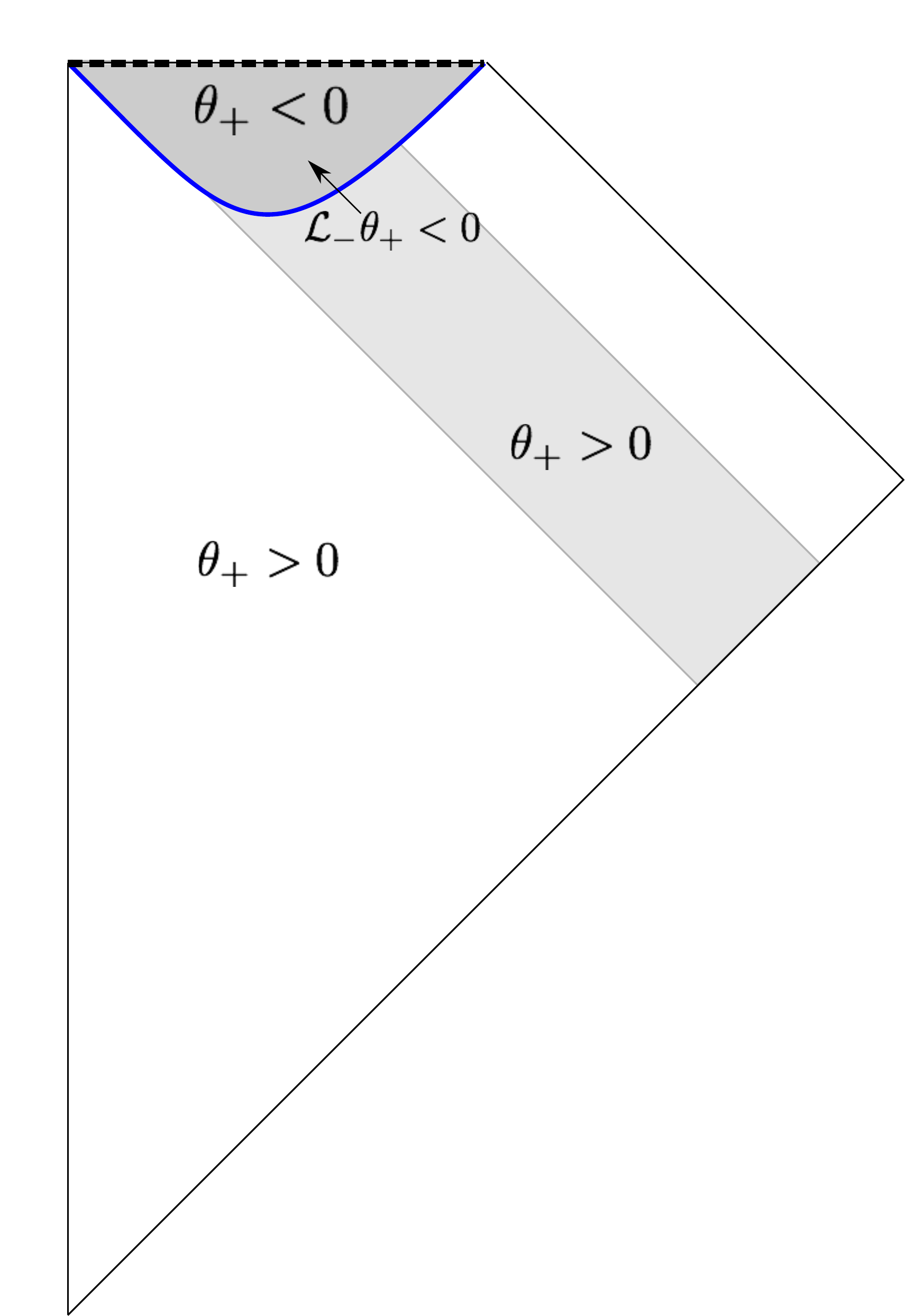}
  \includegraphics[scale=0.3]{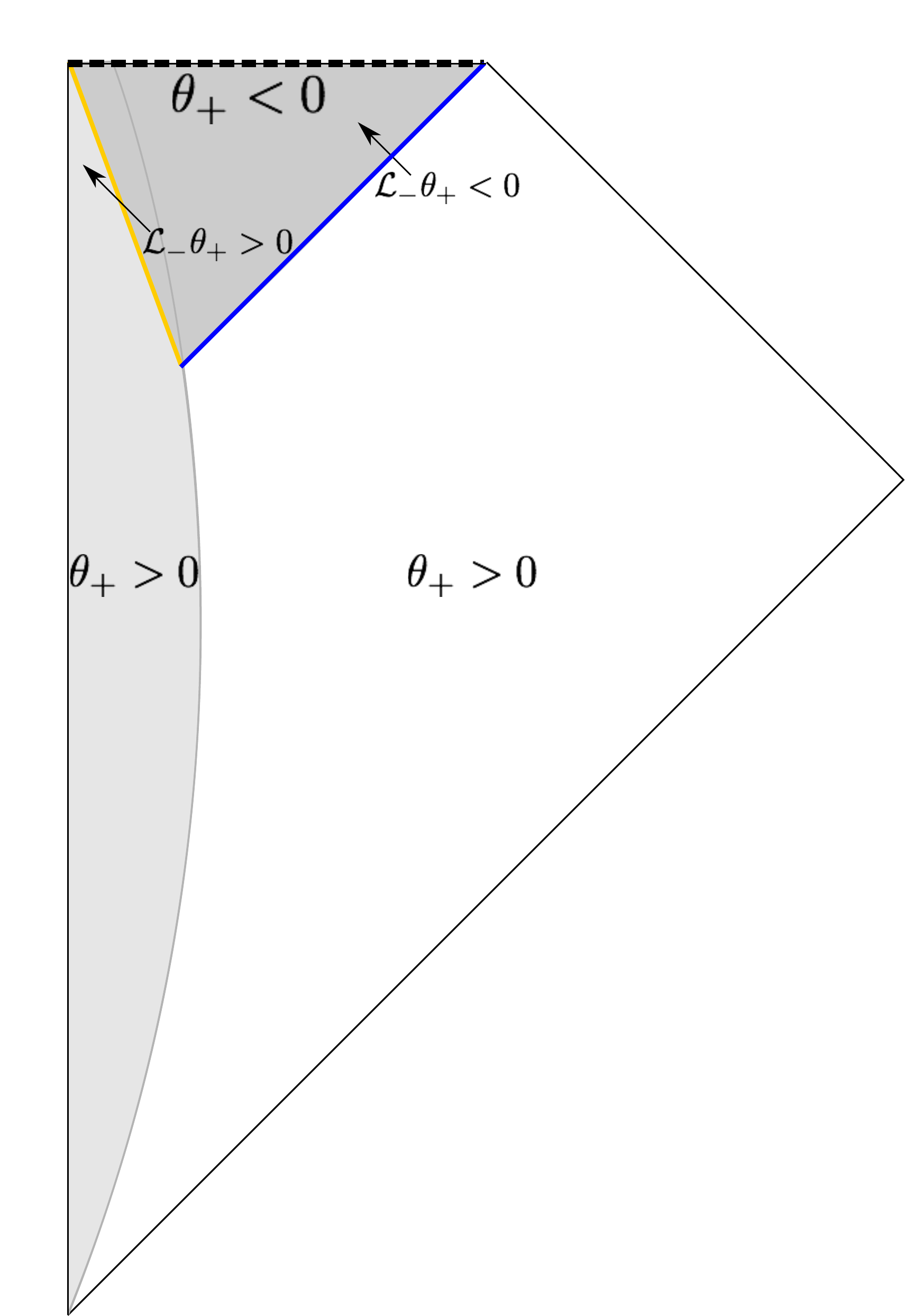}
  \includegraphics[scale=0.3]{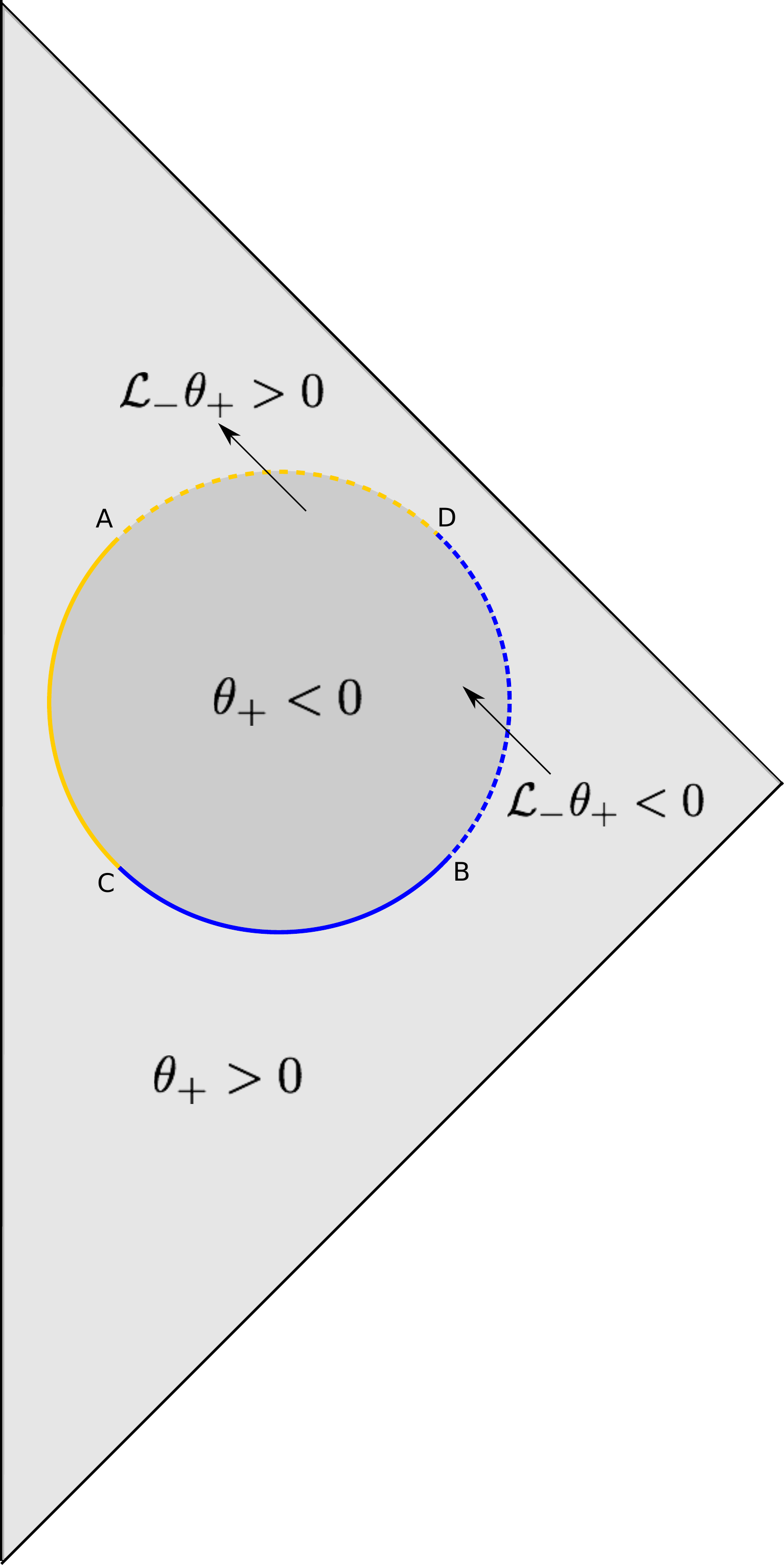}
  \caption{Penrose-Carter diagrams for Vaidya null dust collapse (left), Oppenheimer-Snyder homogeneous dust collapse (centre) and Roman-Bergmann closed trapping horizons \cite{Roman:1983zza} (right). The outer horizon is represented in blue while the inner one is shown in yellow. These are defined from \cite{Hayward:1993wb} using the Lie derivative, along the ingoing null direction, of the expansion of outgoing null geodesic congruence $\theta_+$: $\mathcal{L}_-\theta_+<0$ for outer trapping horizons, $\mathcal{L}_-\theta_+>0$ for inner trapping ones. The horizons are drawn as a solid line when the NEC is satisfied, and as a dashed line when it is violated.}
  \label{fig:Roman_Bergmann}
\end{figure}

The idea of closed apparent/trapping horizons was studied in \cite{Roman:1983zza}, where it was given the general form of Fig.~\ref{fig:Roman_Bergmann} (right panel). This horizon is null at four points $A$, $B$, $C$ and $D$. In the classical diagrams, the apparent horizon is usually spacelike, i.e.~only a portion of $CB$, where $B$ is the point at which the black hole becomes isolated and the apparent horizon becomes indistinguishable from the event horizon. In some situations, one can also have a timelike inner horizon, i.e.~a portion of $CA$ \cite{Booth:2005ng,Helou:2016xyu}. The reason why a classical black hole cannot produce a horizon on portion $ADB$ is the following: when the Null Energy Condition (NEC) is satisfied, an outer horizon is spacelike while an inner horizon is timelike (see Theorem 2 of \cite{Hayward:1993wb}, as well as \cite{Hayward:2005ny, Booth:2005ng}).
Therefore we must have a violation of the NEC on portion $ADB$. Considerations of this portion seldom appear in the literature, although it is inherent to the widely discussed Hawking radiation of the outer horizon, which produces a timelike horizon of type $BD$ (see Fig.~2 of \cite{Hiscock:1980ze}). Having a spacelike inner horizon of the type $AD$ is even less considered (see however Fig.~2 of \cite{Bardeen:2014uaa}), but it is a way to avoid the conclusions of the Penrose singularity theorem \cite{Penrose:1965} by violating the NEC. 

We want to stress here, in accordance with \cite{Frolov:1979, Roman:1983zza, Hayward:2005gi, Bardeen:2014uaa}, that one should not \emph{a priori} discard any of the above behaviours for the trapping horizon. As we still do not know what happens (beyond General Relativity) when the inner horizon reaches the centre of the configuration, or at the end of black hole evaporation, we think it is worth investigating these possibilities, which display a very different phenomenology from the usual classical and semi-classical picture.

In the following, we will investigate the conditions needed in order to get a regular spacetime with closed trapping horizons of the Roman-Bergmann type. In the remainder of this section, we will study the phenomenology of some examples of this general class, that have already been proposed in the literature.

\subsection{Existence of singularity-free spacetimes with closed trapping horizons}

Let us study the general conditions for the existence of a non singular spacetime containing closed trapping horizons. We will work in advanced Eddington-Finkelstein coordinates, in which the metric reads

\begin{equation}\label{metric1}
ds^2=-F(v,R)dv^2+2dvdR+R^2d\Omega^2 \ ,
\end{equation}
where $R$ is the areal radius, $d\Omega^2=d\theta^2+\sin^2 \theta d\phi^2$, and $F$ is a function of $v$ and $R$ not specified yet. This is not the most general spherically symmetric metric, whose expression will be used later (Eq.\eqref{metric_general}). However, as Eq.\eqref{psi_horizons} will illustrate, the additional degree of freedom of the general metric does not affect the shape of the horizons given by the metric \eqref{metric1} used in the present section. 

We will assume that $F$ can be written in the following way
\begin{equation}\label{Fpolynomial}
F(v,R) \equiv 1-\frac{2M(v,R)}{R}=1-2m(v)\frac{R^{\alpha-1}+a_{\alpha-2}(v)R^{\alpha-2}+\cdots+a_1(v)R+a_0(v)}{R^{\alpha}+b_{\alpha-1}(v)R^{\alpha-1}+\cdots+b_1(v)R+b_0(v)}, \hspace{0.5cm} b_0(v)\neq 0 \ .
\end{equation}
Assuming that $F-1$ can be written as a ratio of polynomial functions, this is the most general form one can use to recover Schwarzschild's limit when $R \rightarrow +\infty$. The function $m(v)$ plays the role of the Misner-Sharp mass for an observer at infinity. 

\subsubsection{Conditions for the existence of closed trapping horizons}

The existence of closed trapping horizons requires the presence of two horizons, i.e.\ of one marginally outer and one marginally inner trapped surfaces, whose coordinates $R_1(v)$ and $R_2(v)$ match for at least two different values of $v$.

As mentioned in the Introduction, the locus of the marginally trapped surfaces is defined via the expansion of null outgoing geodesic congruence
\begin{equation}
\theta_+ \equiv h^{ab}\nabla_a k_b=0 \ ,
\end{equation}
where $h_{ab}$ is the induced metric on the 2-spheres of symmetry and $k_b$ an outgoing radial null vector.

For the metric \eqref{metric1}, the expansion is $\theta_+ =\frac{F}{R}$ and thus the locations of the horizons $R(v)$ are defined by

\begin{equation}
\theta_+=0 \Leftrightarrow F(v,R(v))=0 \ .
\end{equation}
Since the existence of a closed trapped region requires the presence of two horizons, the equation $F(v,R(v))=0$ should thus be at least of degree 2 in $R$. In this minimal case of degree 2, one has

\begin{equation}
F(v,R)=1-2m(v)\frac{R+a_0(v)}{R^2+b_1(v)R+b_0(v)} \ ,
\end{equation}
and 

\begin{equation}
\begin{aligned}
F(v,R)=0 & \Leftrightarrow R^2+(b_1(v)-2m(v))R+b_0(v)-2m(v)a_0(v)=0 \\
& \Leftrightarrow 
\begin{array}{ll}
        R(v)=\frac{2m(v)-b_1(v)\pm\sqrt{(b_1(v)-2m(v))^2-4(b_0(v)-2m(v)a_0(v))}}{2} \ .
\end{array}      
\end{aligned}
\end{equation}
Another condition is that there must exist two different $v$ at which $R_1=R_2$, so that the trapping horizons be closed. It is thus entirely possible to choose $a_0(v)$, $b_1(v)$ and $b_2(v)$ in order to construct closed trapping horizons. 
However such a spacetime cannot be singularity-free, as we shall now see.

\subsubsection{Conditions for the absence of singularities}

In order to investigate the presence of a singularity in our spacetime, we need to verify that no curvature scalar diverges at one point of spacetime. We will thus compute the Ricci and Kretschmann scalars, which will give us constraints on the parameter $\alpha$ in Eq.\eqref{Fpolynomial}. They read

\begin{equation}
\left\{
\begin{array}{ll}
        \mathcal{R}=g_{\mu\nu}R^{\mu\nu}=-\frac{R^{2}
\frac{\partial^2\,F}{\partial R^2} + 4 \, R \frac{\partial\,F}{\partial
R} + 2 \, F\left(v, R\right) - 2}{R^{2}} \\
        \mathcal{K}=R_{\mu\nu\rho\sigma}R^{\mu\nu\rho\sigma}=\frac{R^{4} \left(\frac{\partial^2\,F}{\partial
R^2}\right)^{2} + 4 \, R^{2} \left(\frac{\partial\,F}{\partial
R}\right)^{2} + 4 \, F\left(v, R\right)^{2} - 8 \, F\left(v, R\right) +
4}{R^{4}} \ .
         \end{array}
\right.        
\end{equation}
Let us focus on the Ricci scalar first. One  has to ensure that the expression 

\begin{equation*}
 \frac{R^{2}}{2}\frac{\partial^2\,F}{\partial R^2}+2 \, R \frac{\partial\,F}{\partial R} + F\left(v, R\right) -1 \ ,
\end{equation*}
is at least of degree $2$ in $R$ to avoid the presence of a singularity.
\par First of all one notices that $b_0(v) \neq 0$ so that $F$ does not diverge when $R \rightarrow 0$. This implies that $\frac{\partial^2 F}{\partial R^2}$ will contain no divergence, and  $\frac{R^2}{2}\frac{\partial^2 F}{\partial R^2}$ will be at least of degree 2.
\par Then, one can show that

\begin{equation}
2 \, R \frac{\partial\,F}{\partial R} + F\left(v, R\right) -1=\frac{-2m(v)}{\left( R^{\alpha}+\cdots+b_0  \right)^2} \left[\cdots +R^2 \left( 5a_2b_0+a_1b_1-3a_0b_2 \right) +R \left(3a_1b_0-b_1a_0 \right)+a_0b_0 \right] \ ,
\end{equation} 
where the dots denote higher order terms in $R$. Since $b_0 \neq 0$, one must have $a_0=0$ and $a_1=0$ so that the expression in brackets be at least of degree 2. This means that the first non-zero coefficient must be $a_2$, which implies

\begin{equation}
\boxed{\alpha \geq 3} \ .
\end{equation}
A similar reasoning with the Kretschmann scalar leads to the same result, $\alpha \geq 3$. 

\subsubsection{Minimal form of F}

This draws us to the conclusion that the simplest form of $F$ describing a spacetime without singularities and containing closed trapping horizons, as well as allowing to recover Schwarzschild's solution when $R\rightarrow +\infty$, will have the general, minimal form

\begin{equation}
F(v,R)=1-2m(v)\frac{R^2}{R^3+b_2(v)R^2+b_1(v)R+b_0(v)} \ .
\end{equation}
Since we are only interested in the asymptotic behaviours, we can choose for simplicity $b_1(v)=b_2(v)=0$. Then, by writing $b_0(v)$ as $b_0(v)=2m(v)b^2$, we get

\begin{equation}
F(v,R)=1-\frac{2m(v)R^2}{R^3+2m(v)b(v)^2} \ ,
\label{metric_Hayward}
\end{equation}
where we recover Hayward's metric \cite{Hayward:2005gi} when we set $b(v)=b=\text{cst}$. This metric has the interesting property of exhibiting a de Sitter limit when $R \rightarrow 0$, on top of the Schwarzschild limit when $R \rightarrow \infty$. The constant parameter $b$ plays the role of a de Sitter radius, and is interpreted as a Planckian cutoff \cite{Hayward:2005gi, Bardeen:2014uaa}.

\subsection{Examples of closed trapping horizons}

For now, we have argued that the form of $F$ given by Eq.\eqref{metric_Hayward} is the most simple way of building a singularity-free spacetime with closed trapping horizons while recovering Schwarzschild and de Sitter limits (provided $b(v)=\mbox{cst}$ for the latter one). Let us then get more specific and obtain the coordinates of the horizons from Eq.\eqref{metric_Hayward}, before entering the details of some specific models.

\subsubsection{Obtaining the horizons}

The location of the  horizons is by definition
\begin{equation}
\theta_+=0 \Leftrightarrow F(v,R(v))=0 \ ,
\end{equation}
which, with our expression \eqref{metric_Hayward} of $F$, boils down to a polynomial equation in $R$

\begin{equation}
 R^3 -2m R^2+2m b^2=0 \ .
 \label{eq_polynom1}
\end{equation}  
Using Cardan's method, one gets the discriminant $\Delta= 4m^2b^2(16m^2 - 27b^2)$. The equation admits at least two distinct real solutions if $\Delta>0$, and two degenerate real solutions if $\Delta=0$. One has

\begin{equation}
\Delta \geq 0 \Leftrightarrow m \geq \frac{3\sqrt{3}}{4}b \ .
\end{equation}
The starting point and endpoint of the trapping horizons in a $(R,v)$ diagram are thus defined by $m = \frac{3\sqrt{3}}{4}b$.
Provided $m \geq \frac{3\sqrt{3}}{4}b$, one finally gets three solutions for a given value of $v$

\begin{align}
 R_j = \frac{4m}{3} \cos\left(\frac{1}{3}\arccos\left (1 - \frac{27 b^2}{8m^2} \right)+\frac{2(j-1)\pi}{3}\right) + \frac{2m}{3} \ , \hspace{0.3cm} j=1, 2, 3 \ .
\end{align}
$R_1$ and $R_3$ are the only positive solutions, describing the outer and inner trapping horizons respectively. In the case where $b \ll m$ (e.g.\ when $b$ is a Planckian cutoff), expanding these solutions in terms of $b/m$ leads to
\begin{equation}\label{Rdevelopment}
\begin{aligned}
\begin{cases}
 R_1= 2m - \frac{b^2}{2m} + o\left(\frac{b^2}{m} \right) \ , \\
 R_3=  b + \frac{b^2}{4m}+o\left(\frac{b^2}{m} \right) \ .
\end{cases}
\end{aligned}
\end{equation}

\subsubsection{Hayward-like model}

\begin{figure}[h]
\centering
\begin{subfigure}{.5\textwidth}
  \centering
  \includegraphics[scale=0.54]{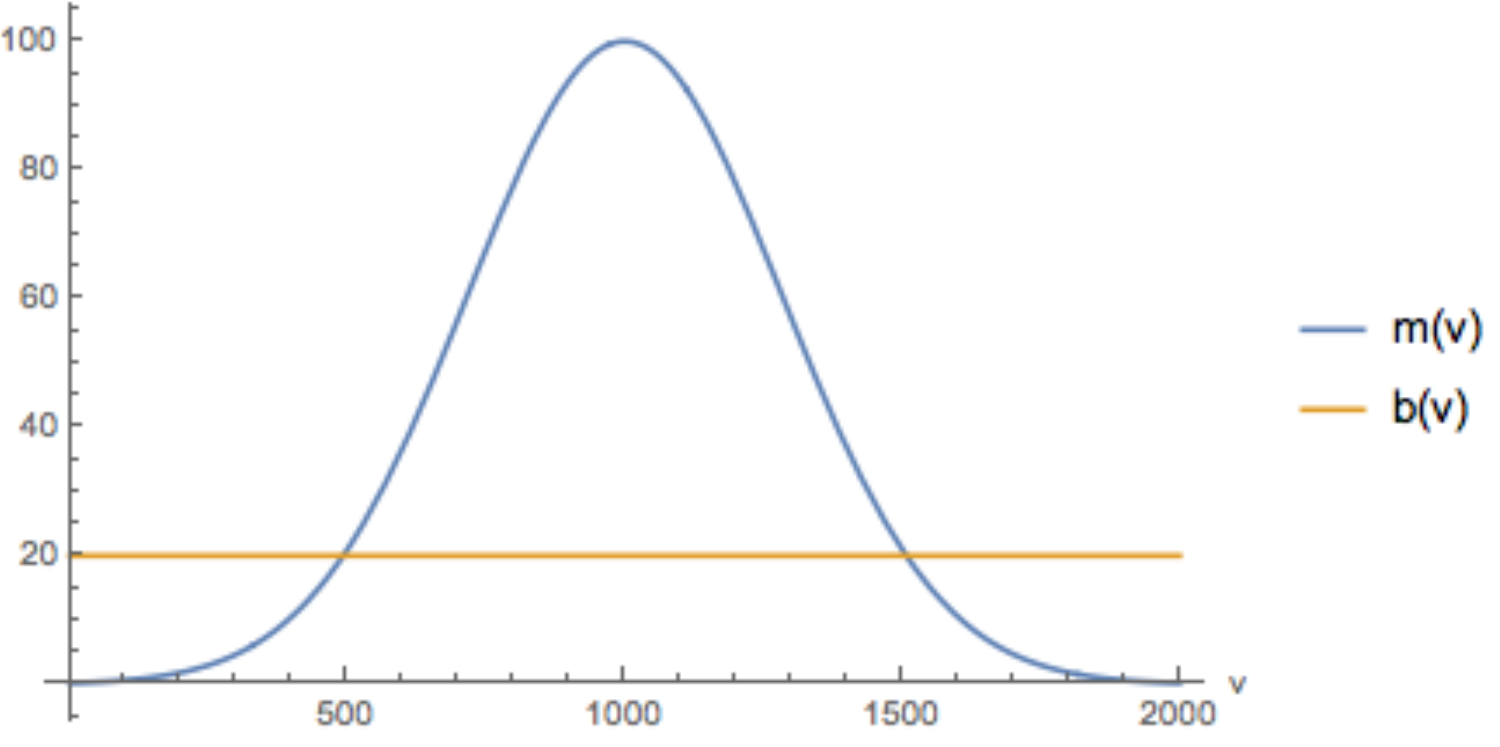}
  \caption{\footnotesize Mass function}
  \label{fig:hay1}
\end{subfigure}%
\begin{subfigure}{.4\textwidth}
  \centering
  \includegraphics[scale=0.5]{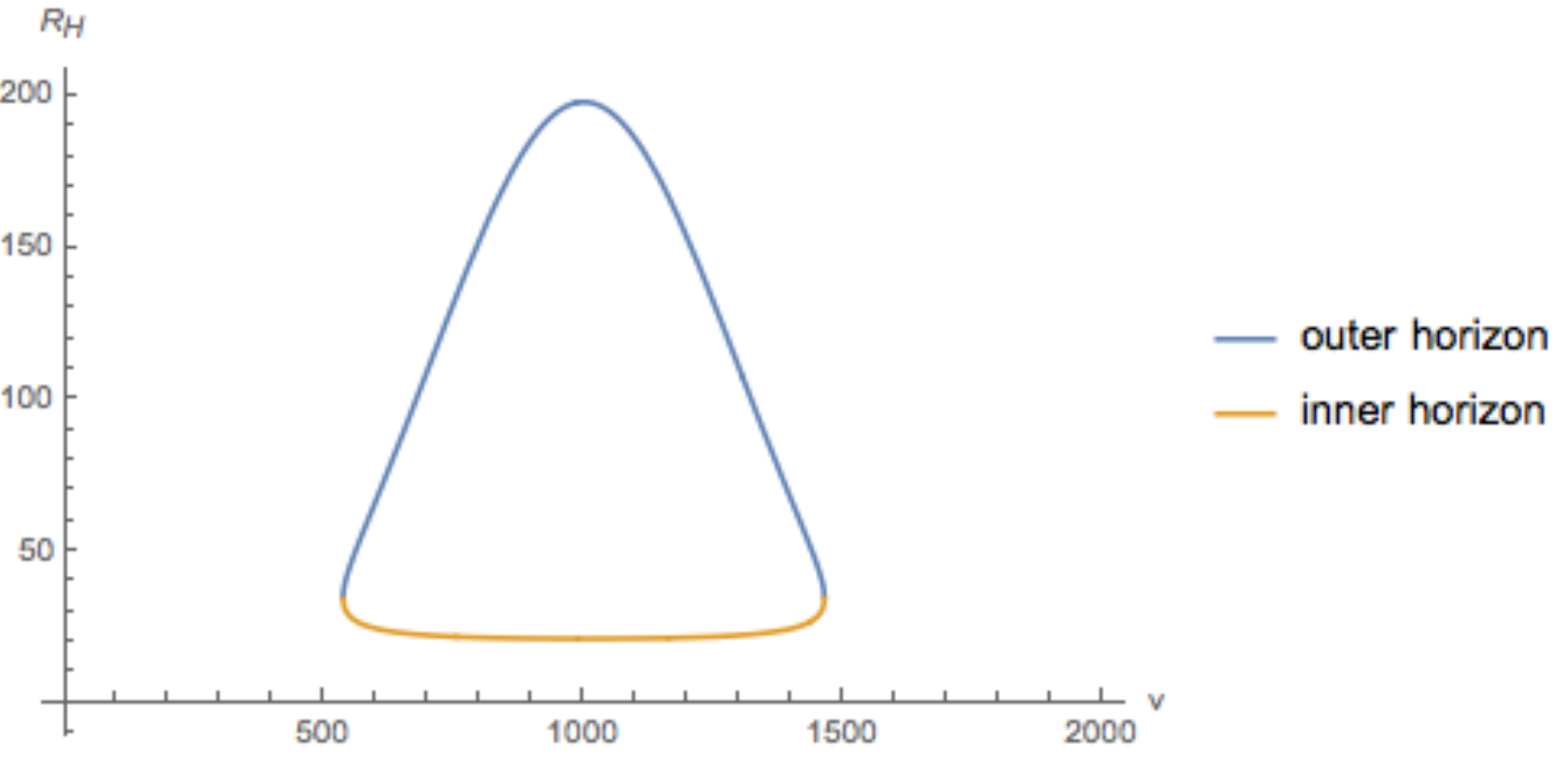}
  \caption{\footnotesize Outer and inner apparent horizons}
  \label{fig:hay2}
\end{subfigure}
\caption{\footnotesize Hayward-like model}
\label{fig:hay}
\end{figure}

Hayward presented in \cite{Hayward:2005gi} a simple model describing the formation and evaporation of a trapped region, relying on the form \eqref{metric_Hayward} of the metric with a constant Planckian cutoff $b$. He chose a symmetric function $m(v)$ containing a plateau which describes similarly the formation and the evaporation phases. We have here used the following form for $m(v)$ and $b$

\begin{align}
 m(v) &=R_0 \exp{\left(-\frac{(v-v_0)^2}{\sigma^2}\right)} \ , \\
 b &=\frac{R_0}{5}  \ ,
\end{align} 
which is plotted on Fig.~\ref{fig:hay1} and where $R_0=100$, $v_0=1000$, $\sigma=400$. Here we chose a macroscopic value for $b$ solely for pedagogical reasons, so that the inner horizon be distinguishable from the horizontal axis on Fig.~\ref{fig:hay2}. This model, although it displays closed trapping horizons and no singularity, suffers from certain limitations in its physical interpretation. 

First of all, let us consider the NEC along the ingoing radial null direction $l^{\mu}$:
\begin{equation}
T_{\mu \nu}l^\mu l^\nu =-\frac{1}{32\pi R} \frac{\partial F}{\partial v} =\frac{1}{16\pi} \frac{m'(v)R^4}{(R^3+2m(v)b^2)^2} \geq 0 \ ,
\end{equation}
where $T_{\mu \nu}$ is the energy-momentum tensor. We see that the NEC is violated when $m'(v)<0$, which happens along lines of constant $v\geq v_0$, $v_0$ being the time when the outer horizon starts shrinking. This is problematic, since it would imply a violation of the NEC in regions arbitrarily far from the collapsed body (e.g. $v=cst$, $R\rightarrow \infty$). In Section \ref{section_energy}, we show that this limitation is inherent to the black hole models asymptotically constructed out of ingoing Vaidya shells. 

Another limitation in the physical interpretation is the symmetry in the outer trapping horizon growing and shrinking. The increase in horizon radius physically comes from the inflow of matter or radiation into the trapped region, while its decrease must come from Hawking radiation. These two effects have no reasons to show the same scaling, which they do in Hayward's model (see Fig.~\ref{fig:hay2}).

Moreover, the reason why the inner trapping horizon is quantum mechanically held at a fixed distance from the centre is not clear, and this feature appears to be quite artificial. Lastly, 
as noticed in \cite{DeLorenzo:2014pta}, this model does not allow for a time delay between the centre of the cloud and infinity since $F \rightarrow 1$ when $R \rightarrow 0$ as well as $R \rightarrow +\infty$.

We will call Hayward-like models those which exhibit symmetric phases of formation and evaporation whilst their inner horizon's radius remains at a Planckianf distance from the centre $R=0$.

\subsubsection{Frolov's model}

\begin{figure}[h]
\centering
\begin{subfigure}{.5\textwidth}
  \centering
  \includegraphics[scale=0.54]{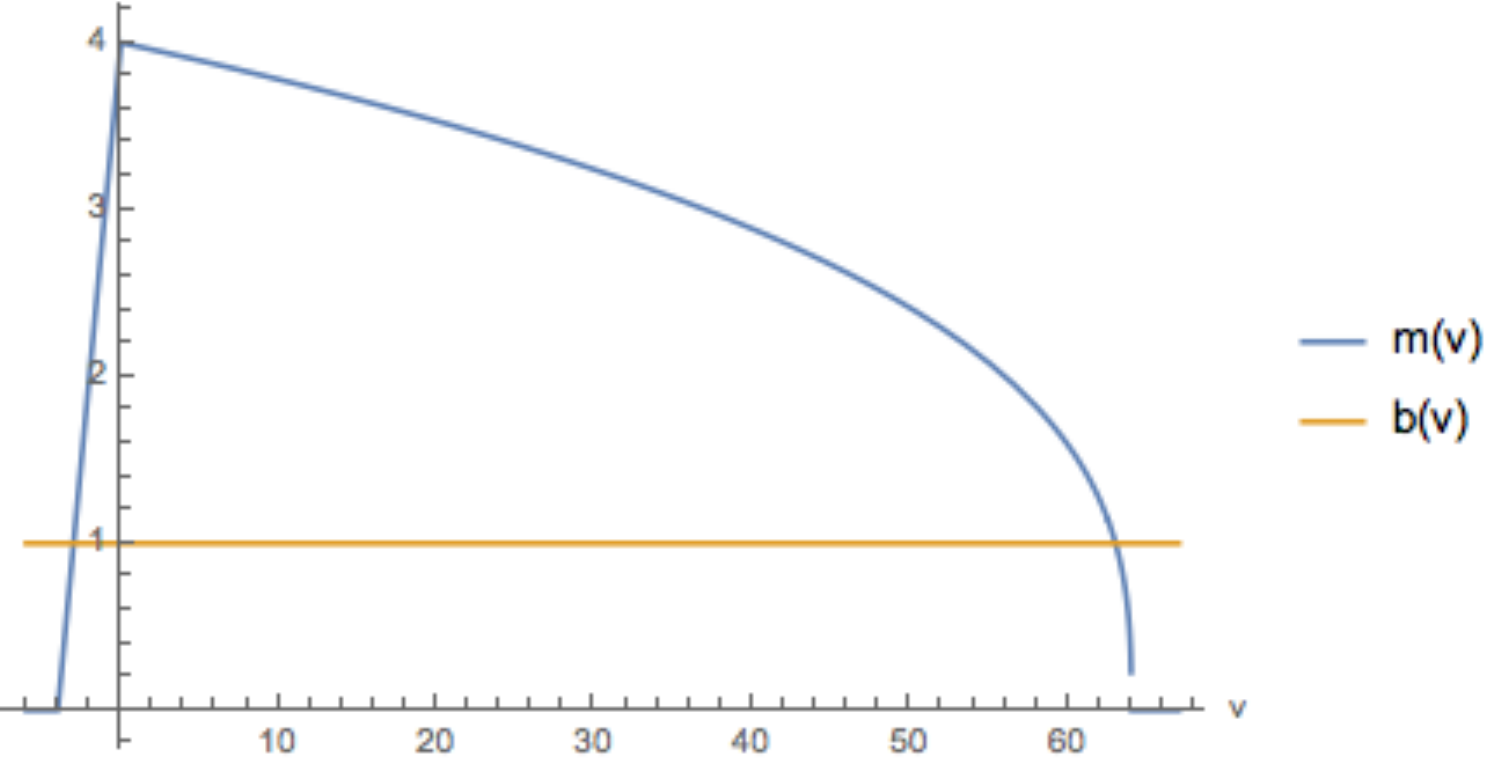}
  \caption{\footnotesize Mass function}
  \label{fig:frol1}
\end{subfigure}%
\begin{subfigure}{.4\textwidth}
  \centering
  \includegraphics[scale=0.5]{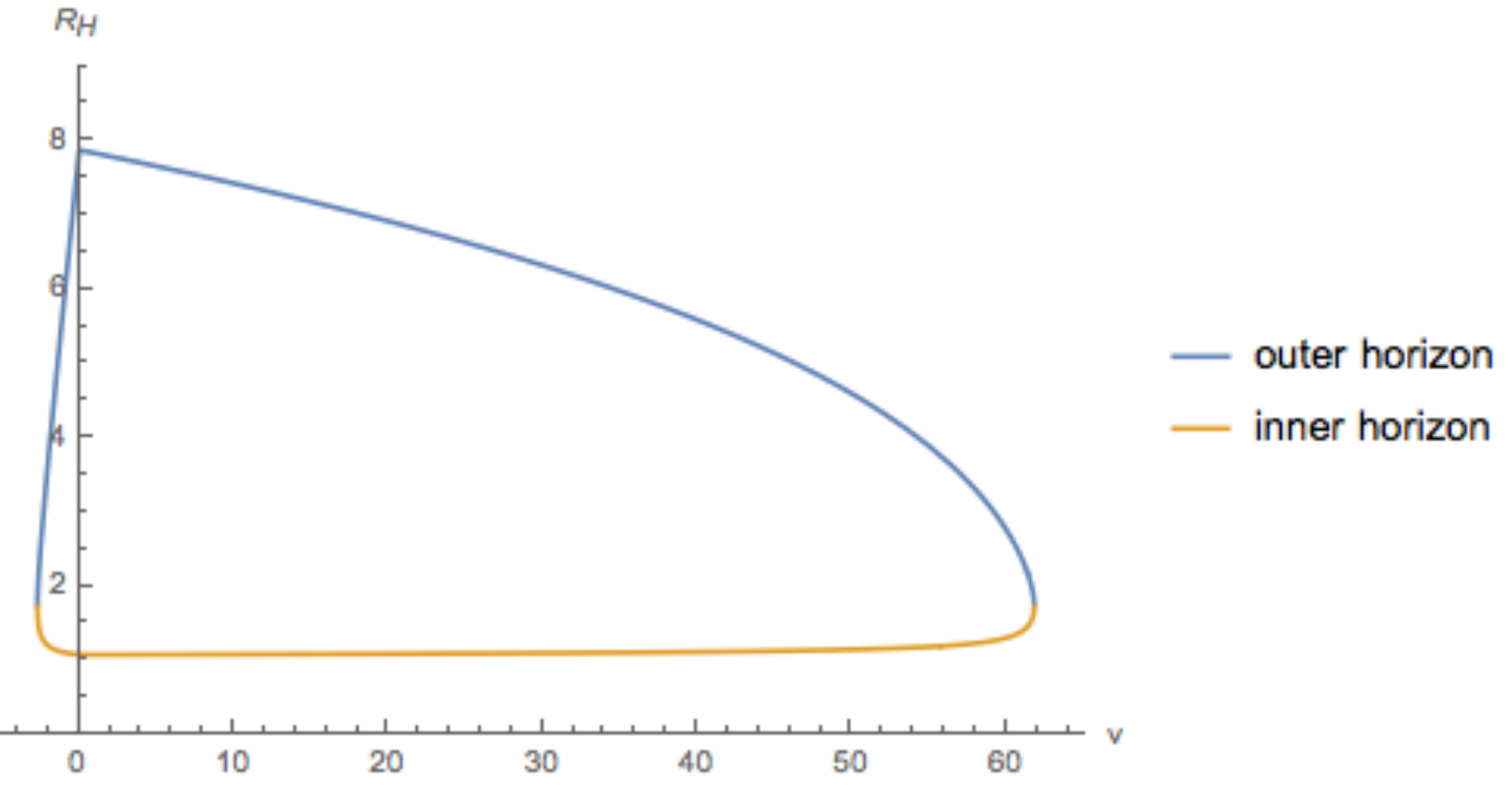}
  \caption{\footnotesize Outer and inner apparent horizons}
  \label{fig:frol2}
\end{subfigure}
\caption{Frolov's model}
\label{fig:frol}
\end{figure}

Frolov's construction \cite{Frolov:2014wja} aims at modelling the Hawking evaporation, and thus introduces a dissymetry between the formation and the evaporation phases. $F$ has the same form as in Hayward's model, but here the mass function is defined by parts

\begin{equation}
\left\{
\begin{array}{llll}
       -\infty<v<v_0 &: m(v)=0 \ , \\
       v_0<v<0 &: m(v)/b = (m_0/b)^3 + v/b  \ , \\
        0<v<v_1 &: (m(v)/b)^3=(m_0/b)^3-v/b \ , \\
        v_1<v<+\infty &: m(v)=0 \ ,
          \end{array}
\right.        
\end{equation}
where $v$, $m(v)$ and $m_0=4$ are expressed in units of $b$. The form of $m(v)$ during the evaporation phase ($0<v<v_1$) is chosen so that one recovers the correct scaling for the mass loss $\dot{m}$ due to Hawking radiation, i.e.
\begin{equation}
 \dot{m}\sim  -C\left(\frac{m_{Pl}}{m}\right)^2 \ ,
\end{equation}
where $C$ is a coefficient depending on the details of the emitted particles, and $m_{Pl}$ is the Planck mass. This gives a more realistic description of the evaporation process than the symmetric model of Hayward.
The obtained shapes for the parameter functions and for the horizons are shown on Fig.~\ref{fig:frol}.

However, this model still displays some important limitations from the point of view of the physical interpretation: the violation of the NEC at infinity when the outer trapping horizon starts shrinking, the constancy of the inner trapping horizon radius, and the absence of time delay between the centre and infinity.

\subsubsection{Bardeen-like model}

In the two previous models, we have noticed that the inner horizon almost stays at a constant and small radius $R$. This comes in fact directly from the expansion \eqref{Rdevelopment}, which implies that the inner horizon radius is essentially given by the constant Planckian cutoff $b$.

\begin{figure}[h]
\centering
\begin{subfigure}{.5\textwidth}
  \centering
  \includegraphics[scale=0.55]{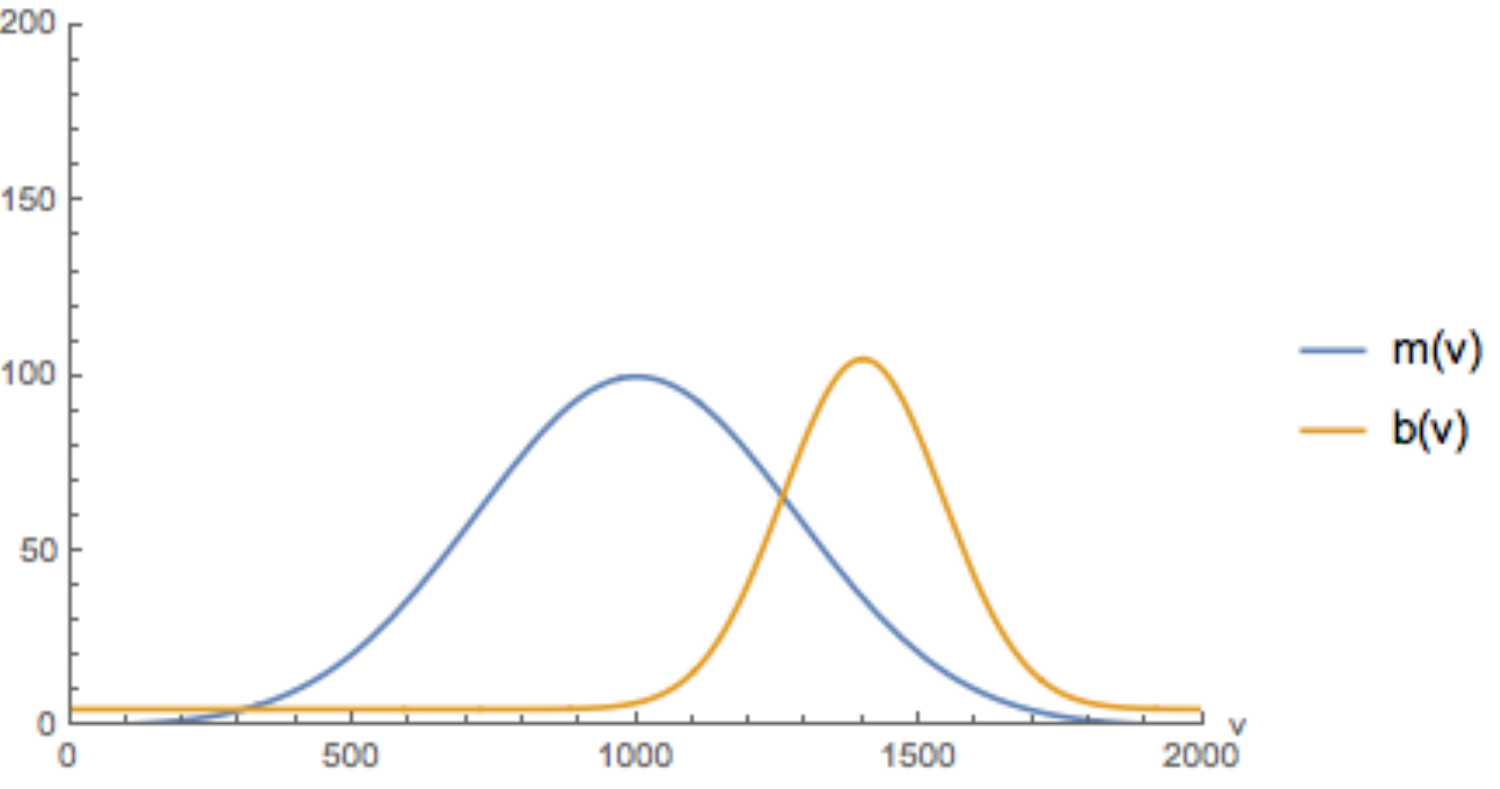}
  \caption{\footnotesize Mass function}
  \label{fig:bard1}
\end{subfigure}%
\begin{subfigure}{.4\textwidth}
  \centering
    \includegraphics[scale=0.5]{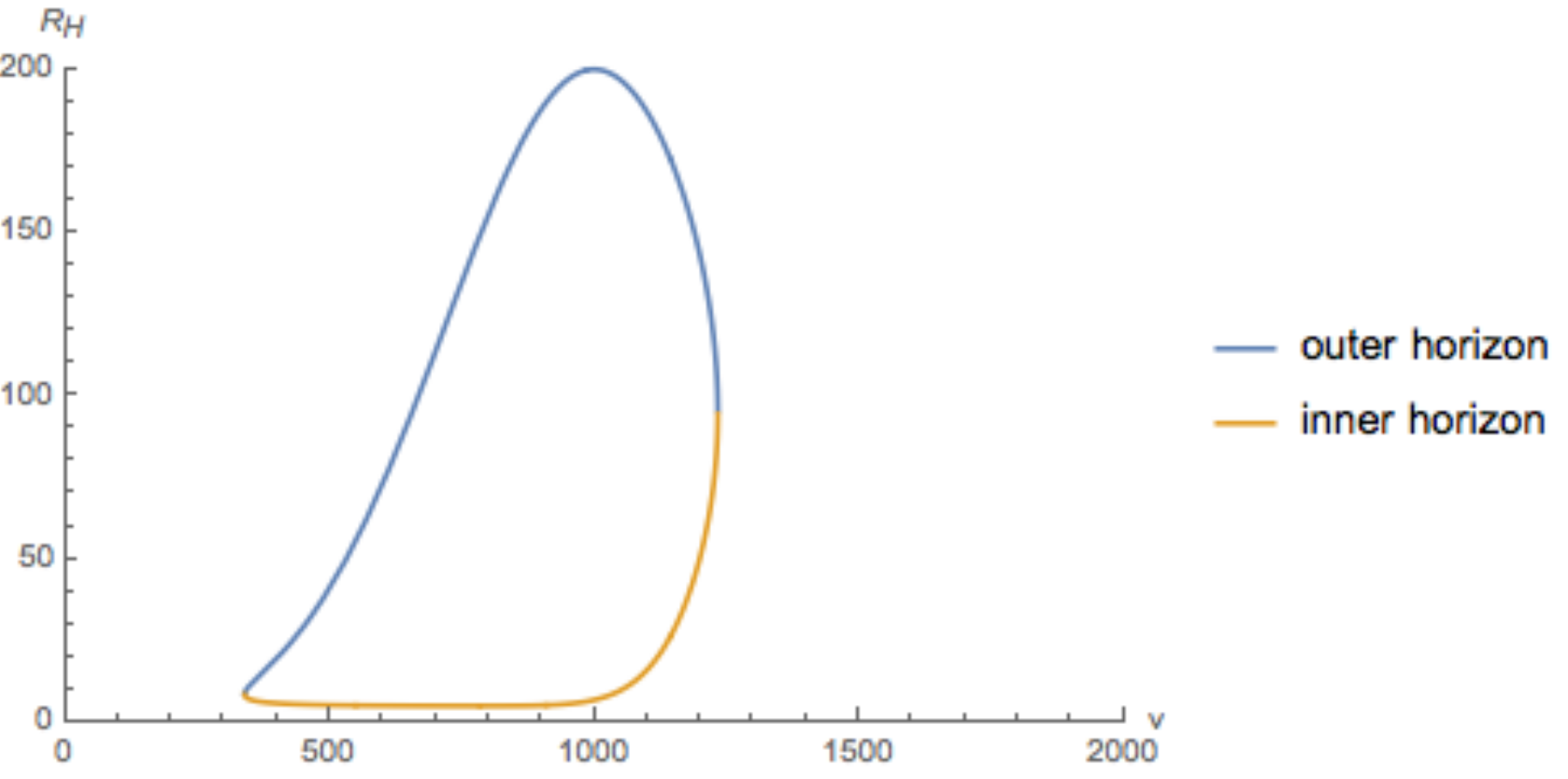}
  \caption{\footnotesize Outer and inner apparent horizons}
  \label{fig:bard2}
\end{subfigure}
\caption{\footnotesize Bardeen-like model}
\label{fig:bard}
\end{figure}

However, one of Bardeen's main points in \cite{Bardeen:2014uaa} consists in giving a dynamics to the inner horizon. More precisely, Bardeen argues that some Hawking pairs will be created at the inner horizon, which will begin to grow due to Bousso's covariant entropy bound \cite{Bousso:1999xy} and finally reach the outer horizon at macroscopic scales. We will thus call Bardeen-like models those which exhibit such a property of the inner horizon (Fig.~\ref{fig:bard2}). We have tried to explicitly recover this model with the following parameter functions, plotted on Fig.~\ref{fig:bard1} 
\begin{align}
 m(v) &=R_0 \exp{\left(-\frac{(v-v_0)^2}{\sigma^2}\right)} \ , \\
 b(v) &= R_0 \exp{\left(-\frac{(v-v_0')^2}{\sigma'^2}\right)}+b_0 \ ,
\end{align} 
where $R_0=100$, $v_0=1000$, $v_0'=800$, $\sigma=400$,  $\sigma'=200$, $b_0=5$.

Once again, this model possesses important limitations in its physical interpretation: the NEC is violated in a non-compact region as soon as the outer trapping horizon starts shrinking, and the time-delay between the centre and an asymptotic clock is absent.

Another model of non-singular black hole, based on the metric \eqref{metric1} with a function $F(v,R)$ given by Eq.\eqref{metric_Hayward}, and known as Planck stars, is free from some of these limitations: the NEC is violated in a compact neighbourhood of the source and the time-delay in the core is present \cite{DeLorenzo:2014pta}. However, this model is static, with fixed values for the radii of the outer and inner trapping horizons, and therefore cannot describe the dynamics of the formation and evaporation of a closed trapped region.

Before presenting in Section \ref{section_energy} our attempt to answer the aforementioned limitations of the models that are found in the literature, let us investigate in more details some properties shared by all these models. That is the topic of Section \ref{section_globalprop}.

\section{Behaviour of null geodesics in models with closed trapping horizons}

The above models do not possess any event horizon since they are dynamic and aim at describing a trapped region eventually fully evaporated, leaving no region of spacetime causally disconnected from future null infinity. It is nonetheless of interest to study the relevant geodesics of such spacetimes.

 \label{section_globalprop}
\subsection{Null geodesic flow}

The radial null geodesics for metric \eqref{metric1},

\begin{equation}
ds^2=-F(v,R)dv^2+2dvdR+R^2d\Omega^2 \ ,
\end{equation}
are given by

\begin{equation}
ds^2=0 \Leftrightarrow \left\{
\begin{array}{ll}
       dv=0 \ , \\
        \frac{dR}{dv}=\frac{F(v,R)}{2} \ .
          \end{array}
\right.        
 \label{eq_null_geodesics}
\end{equation}
In the case of Minkowski spacetime, $F=1$ and the radial null geodesics are trivial since the lightcone is the same at each point of spacetime. In a $(v,R)$ diagram, ingoing radial null geodesics are $v=\mbox{cst}$ lines while outgoing ones are lines of slope $1/2$; this is the behaviour we will recover far from the trapped region. \\

\begin{figure}[htbp]
 \centering
 \begin{subfigure}{.4\textwidth}
   \hspace{-1cm}
   \includegraphics[scale=0.45]{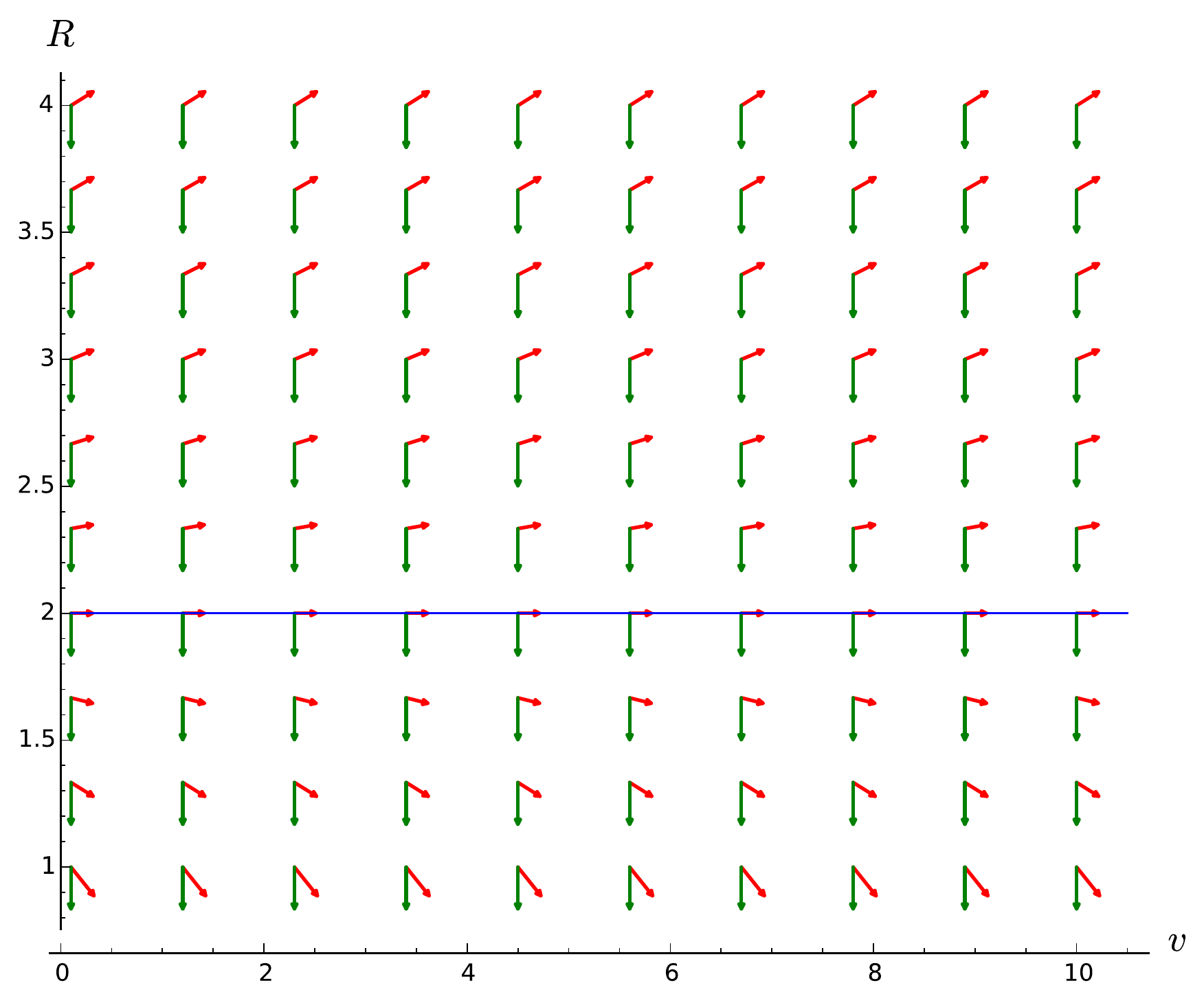}
   \caption{Schwarzschild}
   \label{fig:Schwarzschild_spacetime}
 \end{subfigure}
 \begin{subfigure}{.4\textwidth}
 \hspace{-1cm} \includegraphics[scale=0.75]{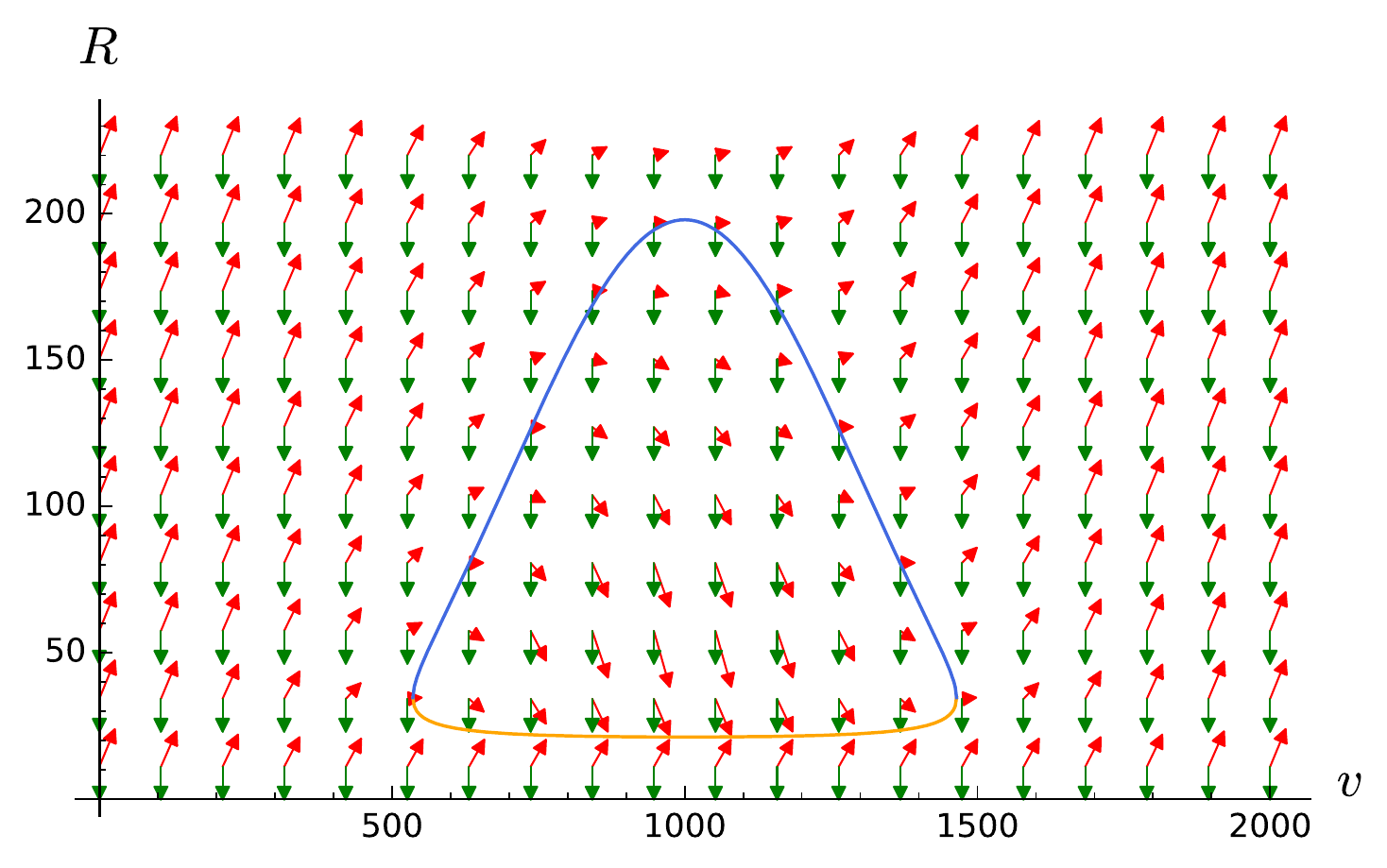}
  \caption{Hayward-like}
   \label{fig:Hayward-like model}
 \end{subfigure}
\caption{ \footnotesize Plot of the outgoing (in red) and ingoing (in green) null vectors depicting the lightcone of Schwarzschild (a) and Hayward-like (b) spacetimes. In Schwarzschild's case there is only one horizon (in blue), which is a null hypersurface for all $v$ hence called \emph{event horizon}. As regards the Hayward-like geometry, the outer (in blue) and inner (in yellow) \emph{trapping horizons} are successively timelike, null and spacelike.}
\end{figure}

When the metric is not trivial, the outgoing radial null geodesics will differ from straight lines in the $(v,R)$ diagram. This can been seen on Fig.~\ref{fig:Schwarzschild_spacetime}, where we have plotted these geodesics for a Schwarzschild metric (with $M=1$). Visualizing the outgoing geodesics reveals the existence of an event horizon: the lightcone prevents any matter or light from the region $R<2$ to escape, and this for all $v$. Therefore there exists a region of spacetime that cannot communicate with $\mathscr{I}^+$, and this region is bounded by an event horizon, by definition. \\

In our models (e.g.\ the Hayward-like model, Fig.~\ref{fig:Hayward-like model}), no event horizon appears. However trapping horizons develop, and are not necessarily tangent to the lightcones. Indeed, trapping horizons are dynamical and can be spacelike, null or timelike \cite{Hayward:1993wb}.

\subsection{Frolov's separatrix and quasi-horizon}

\begin{figure}
 \begin{center}
  \hspace{2.5cm} \includegraphics[scale=0.6]{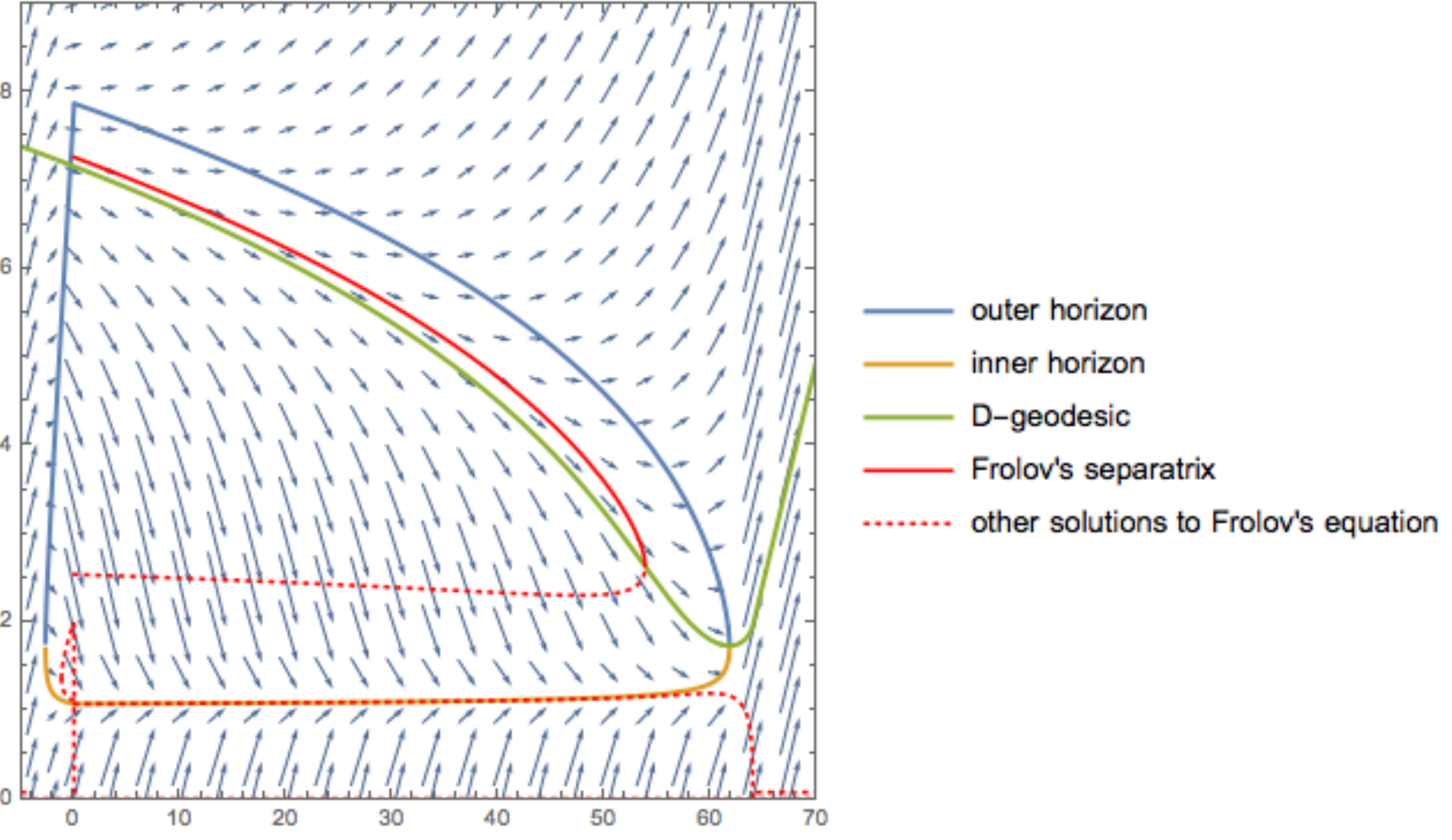}
   \caption{\footnotesize Example of separatrix (in red) and $D$-geodesic (in green) for Frolov's model with $b=1$, $m_0^3=4$. The field of outgoing null vectors (blue arrows) illustrates that the separatrix is traversable by outgoing null or timelike matter, while the $D$-geodesic is not.}
  \label{figure_separatrix}
 \end{center}
\end{figure}

In spite of the absence of a region causally disconnected from future null infinity, there is still a non-trivial behaviour of the null outgoing geodesics due to the trapped region, which is interesting to investigate. In particular, since the apparent horizon can now be timelike and therefore traversable, we may want to look for an alternative surface that would not be traversable from the inside. This surface is easily found to be the null outgoing geodesic which passes through the last trapped sphere, i.e. point $D$ of Fig.~\ref{fig:Roman_Bergmann}. It is the last null outgoing geodesic to leave the trapped region (in terms of time $v$, see Fig.~\ref{figure_separatrix}), and we may call it the $D$-geodesic. This boundary of the no-escape region (a region which has finite lifetime here) is dubbed ``quasi-horizon'' in \cite{Frolov:2014jva}: it traps all the matter it contains until the final evaporation of the trapped region.

It was also suggested in \cite{Frolov:2014jva} to use the separatrix of the null outgoing radial vector field, defined by the vanishing of $\frac{d^2R}{dv^2}$ for the geodesics of Eq.\eqref{eq_null_geodesics}, $\frac{dR}{dv}=\frac{F}{2}$. This yields
\begin{equation}
 \frac{\partial_v F}{2} + \frac{dR}{dv}\frac{\partial_R F}{2} = \frac{\dot{F}}{2} + \frac{F F'}{4} = 0 \hspace{0.5cm} \Rightarrow \hspace{0.5cm} 2\dot{F} +FF' = 0 \ ,
  \label{eq_separatrix}
\end{equation}
where a dot (resp. prime) denotes a derivative with respect to $v$ (resp. $R$). This surface characterizes the strength of the trapping of light rays inside the trapped region: on one side of the separatrix the light rays are more and more trapped, whereas they are less and less so on the other side. When this surface is a null outgoing geodesic in the trapped region, it is not possible for light rays to cross it from inside to outside, and they are doomed to become evermore trapped. This is the case with the Schwarzschild black hole, where $F=1-2M/R$ and $M$ is a constant. Then Eq.\eqref{eq_separatrix} yields $R=2M$ and the separatrix coincides with the apparent horizon which, in this case, is also an event horizon (and of course also a quasi-horizon). 

However in general, the separatrix is not lightlike but can be timelike, and therefore null outgoing geodesics may traverse it. This is visible on Fig.~\ref{figure_separatrix}, where all the solutions to Eq.\eqref{eq_separatrix} have been plotted in red. In this case it cannot coincide with the $D$-geodesic (or ``quasi-horizon''), and we keep the latter notion to be the relevant one in the study of the region of non-escaping matter and radiation.

In \cite{Frolov:2014jva}, closed trapping horizons are built, with the separatrix and $D$-geodesic taken as synonymous, most certainly because the separatrix is close to being null in this particular case. We nevertheless stress the fact that in general, the two notions are distinct.

\subsection{Relevant null geodesics for closed trapping horizons}

We defined above the $D$-geodesic, called quasi-horizon by Frolov, which enables to divide particles located in the trapped region into two categories: those exiting from this region by the outer horizon, and those exiting by the inner one.

\vspace{0.3cm}

Other regions are relevant for the study of models of closed trapping horizons, and can be defined by using radial null outgoing geodesics going through not only $D$ but also points $A$, $B$, and $C$ of Fig.~\ref{fig:Roman_Bergmann}. The $A$-geodesic goes through the last point (in terms of $v$) at which the inner horizon is null. This curve thus bounds from above the region of spacetime whose content is causally prevented from going into the trapped region.
The $B$-geodesic goes through the first point (in terms of $v$) at which the outer horizon is null. It represents the first geodesic (in terms of $v$) able to escape from the trapped region. 
Finally, it may also be interesting to define the $C$-geodesic as the geodesic going through the point of formation of the two horizons. It divides massless outgoing particles located in the trapped region into two categories: those which entered via the outer horizon, and those which did via the inner horizon. 

\vspace{0.3cm}

These geodesics allow to define (at least) two zones of the spacetime which have a physical significance:\\
(i) all massless outgoing particles of the trapped region which do not exit by the outer horizon  must belong to a zone bounded by the $A$-geodesic and the $D$-geodesic. \\
(ii) all Hawking particles emitted from the outer horizon must belong to a zone bounded by the $B$-geodesic and $A$-geodesic.

The $A$ and $D$-geodesics are plotted below for the three different models. In each model these two geodesics quickly tend towards those of Minkowski's spacetime (slope $1/2$) after the disappearance of the trapped region. They delineate a corridor whose largest version is associated with the Bardeen-like case.

\begin{figure}[!h]
\centering
\includegraphics[scale=0.7]{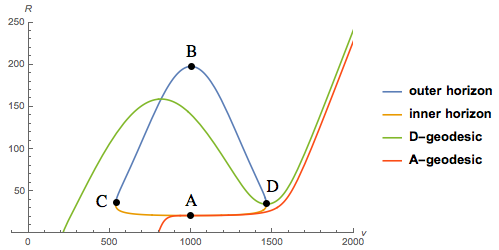}
\caption{\footnotesize Hayward-like model}
\end{figure}
\par

\begin{figure}[!h]
\centering
\includegraphics[scale=0.7]{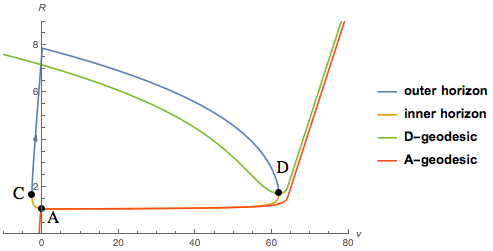}
\caption{\footnotesize Frolov's model. There is no point $B$ since $m(v)$ is not differentiable.}
\end{figure}

\begin{figure}
\centering
\includegraphics[scale=0.7]{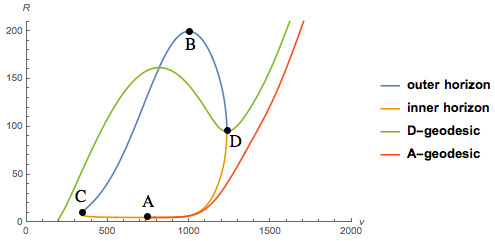}
\caption{\footnotesize Bardeen-like model}
\end{figure}

\newpage

\section{Energy-momentum tensor}
 \label{section_energy}

For now we have been giving a geometry describing the formation and evaporation of closed trapping horizons without singularity, while requiring solely that we recover the appropriate Schwarzschild and de Sitter limits. It is now necessary to study the associated energy content, by solving Einstein's equations, in order to ensure that this content is physical. A first hint can be given by the analysis of the energy conditions, and particularly of the weakest of all: the null energy condition. This will indicate us that an extra degree of freedom is needed in the metric in order to describe all phases of the formation and evaporation of the trapped region. We will then obtain an explicit form of $F$ describing the formation of a trapped region from an ingoing null shell and its evaporation into an outgoing null shell.

\subsection{Form of the energy-momentum tensor}

Let us start with the most general spherically symetric metric in advanced Eddington-Finkelstein coordinates, encoding a new degree of freedom through the function $\psi$:

\begin{equation}\label{metric_general}
ds^2=-F(v,R)\mbox{e}^{2\psi(v,R)}dv^2+2\mbox{e}^{\psi(v,R)}dvdR+R^2d\Omega^2 \ .
\end{equation}
This form will prove to be useful later on, since we will show that no evaporation can occur with a constant $\psi$. By virtue of Einstein's equations, the energy-momentum tensor can be written

\begin{equation} \label{EMT_dv,dR}
T=T_{vv} dv \otimes dv + T_{vR}  (dv \otimes dR+dR \otimes dv)+ T_{RR} dR \otimes dR + T_{\theta \theta} \left(d\theta \otimes d\theta + \sin^2 \theta d\phi \otimes d\phi \right) \ ,
\end{equation}
with

\begin{displaymath}
\left\{
\begin{array}{llll}
8 \pi T_{vv}&=  -\frac{1}{R^2} \left(R F\left(v, R\right) e^{2 \, \psi\left(v, R\right)} \frac{\partial\,F}{\partial R} + F\left(v, R\right)^{2} e^{2 \, \psi\left(v, R\right)} + R e^{\psi\left(v, R\right)} \frac{\partial\,F}{\partial v} - F\left(v, R\right) e^{2 \, \psi\left(v, R\right)} \right) \ , \\
8 \pi T_{vR}&=  \frac{1}{R^2} \left(R e^{\psi\left(v, R\right)} \frac{\partial\,F}{\partial R} + F\left(v, R\right) e^{\psi\left(v, R\right)} - e^{\psi\left(v, R\right)} \right) \ , \\
8 \pi T_{RR}&= \frac{2}{R} \frac{\partial\,\psi}{\partial R} \ , \\
8 \pi T_{\theta \theta}&=  \frac{1}{2} \left[2 \, R^{2} F\left(v, R\right) e^{\psi\left(v, R\right)} \left(\frac{\partial\,\psi}{\partial R}\right)^{2} + 2 \, R^{2} F\left(v, R\right) e^{\psi\left(v, R\right)} \frac{\partial^2\,\psi}{\partial R ^ 2} + R^{2} e^{\psi\left(v, R\right)} \frac{\partial^2\,F}{\partial R ^ 2} + 2 \, R e^{\psi\left(v, R\right)} \frac{\partial\,F}{\partial R} \right. \\ 
& \left. +2 \, R^{2} \frac{\partial^2\,\psi}{\partial v\partial R} + {\left(3 \, R^{2} e^{\psi\left(v, R\right)} \frac{\partial\,F}{\partial R} + 2 \, R F\left(v, R\right) e^{\psi\left(v, R\right)}\right)} \frac{\partial\,\psi}{\partial R}\right] e^{-\psi\left(v, R\right)} \ .
\end{array}
\right.
\end{displaymath}
In the following, we will attempt to compare our energy content to the one of a pure Vaidya spacetime. Indeed, the collapse of an ingoing null shell described by an ingoing Vaidya metric is a natural candidate for the formation phase of the trapped region written in $(v,R)$ coordinates. Moreover, the Hawking radiation associated with the evaporation phase can be described at first order by a flux of outgoing photons, hence the use of an outgoing Vaidya metric. To that purpose, let us write our energy-momentum tensor as follows

\begin{equation} \label{EMT_null}
T=T_{kk} k \otimes k + T_{ll}  l \otimes l + T_{kl} \left( k \otimes l + l \otimes k \right) + T_{\theta \theta} \left(d\theta \otimes d\theta + \sin^2 \theta d\phi \otimes d\phi \right) \ ,
\end{equation}
where $l$ and $k$ are two independent null covectors, respectively ingoing and outgoing. Notice that it is always possible to write the energy-momentum tensor under this form under the assumption of spherical symmetry; the coefficients $T_{kk}$, $T_{ll}$ and $T_{kl}$ solely depend on the non-spherical components of the metric. \\

To get to the form \eqref{EMT_null}, one needs the expressions of the outgoing and ingoing radial null covectors $k$ and $l$

\begin{equation}
\left\{
\begin{array}{ll}
k & =-\frac{F}{2}e^{2\psi}dv+e^{\psi}dR \ , \\
l &=-2dv \ ,
\end{array}
\right.
\end{equation}
where the normalization $k \cdot l=-2$ has been chosen. \\

One then obtains

\begin{equation} \label{dv,dR}
\left\{
\begin{array}{ll}
dv & =-\frac{1}{2}l \ , \\
dR &=e^{-\psi}k-\frac{Fe^{\psi}}{4}l \ .
\end{array}
\right.
\end{equation}
Pluging \eqref{dv,dR} into \eqref{EMT_dv,dR} finally leads to

\begin{equation} \label{EMT_null_comp}
\left\{
\begin{array}{lllll}
8 \pi T_{kk}&=8 \pi T_{RR}e^{-2\psi}=\frac{2 \, e^{-2 \, \psi\left(v, R\right)} \frac{\partial\,\psi}{\partial R}}{R} \ , \\
8 \pi T_{ll}&=8 \pi \left( \frac{T_{vv}}{4}+\frac{Fe^{\psi}}{4}T_{vR}+\frac{F^2e^{2 \psi}}{16} \right)=\frac{F\left(v, R\right)^{2} e^{2 \, \psi\left(v, R\right)} \frac{\partial\,\psi}{\partial R} - 2 \, e^{\psi\left(v, R\right)} \frac{\partial\,F}{\partial v}}{8 \, R} \ , \\
8 \pi T_{kl}&=8 \pi \left(-\frac{e^{-\psi}}{2}T_{vR}-\frac{F}{4}T_{RR} \right)=-\frac{R F\left(v, R\right) \frac{\partial\,\psi}{\partial R} + R \frac{\partial\,F}{\partial R} + F\left(v, R\right) - 1}{2 \, R^{2}} \ .
\end{array}
\right.
\end{equation}

\subsection{NEC violation} 
\label{NEC violation}

As we have already mentioned in Section \ref{section2.1}, models of trapped region with closed horizons require a violation of the null energy condition on the interval $ADB$ of Fig 1. When the NEC is violated, so are all the energy conditions. It is thus of interest to verify that this violation occurs in a region of finite size, i.e.\ that the violation is confined to a compact region of spacetime. \\

Recall that the NEC is expressed as follows: for all null vector $n^\mu$,
\begin{equation} \label{NEC}
T_{\mu \nu}n^\mu n^\nu \geq 0 \ .
\end{equation}

\begin{figure}
\centering
\includegraphics[scale=0.7]{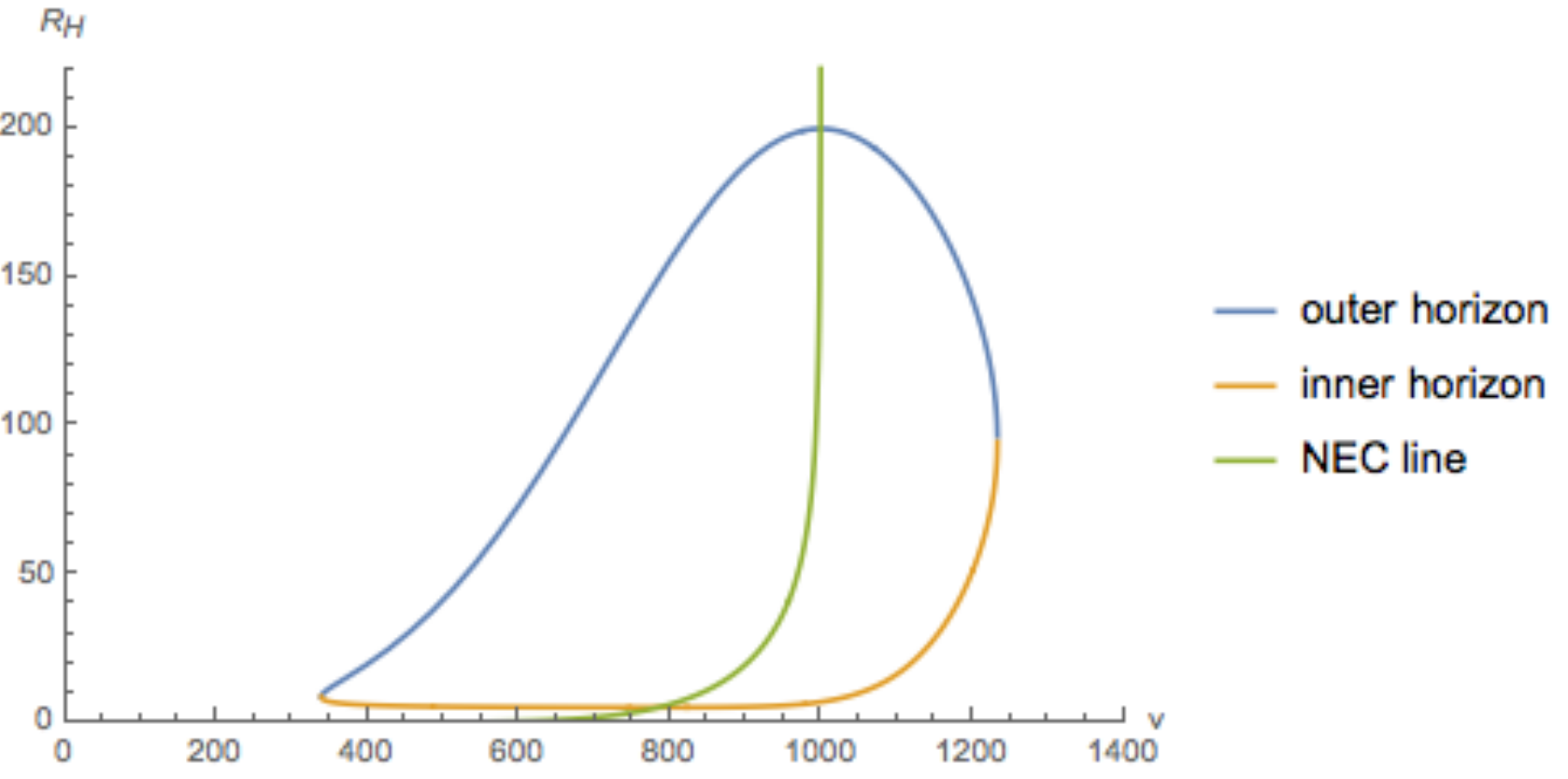}
\caption{\footnotesize NEC violation in Bardeen-like model}
 \label{fig_NEC_violation_Bardeen}
\end{figure}

Using the covectors $l$ and $k$ as long as equations \eqref{EMT_null} and \eqref{EMT_null_comp}, the NEC then reads

\begin{equation} 
\left\{
\begin{array}{ll}
T_{\mu \nu}k^\mu k^\nu & =T_{kk}=\frac{2 \, e^{-2 \, \psi\left(v, R\right)} \frac{\partial\,\psi}{\partial R}}{8 \pi R} \geq 0 \ , \\
T_{\mu \nu}l^\mu l^\nu &=T_{ll}=\frac{F\left(v, R\right)^{2} e^{2 \, \psi\left(v, R\right)} \frac{\partial\,\psi}{\partial R} - 2 \, e^{\psi\left(v, R\right)} \frac{\partial\,F}{\partial v}}{64 \pi R} \geq 0 \ .
\end{array}
\right.
\end{equation}
Let us now focus on the cases of the models developed above, namely Hayward, Frolov and Bardeen-like models. In this case, $\psi(v,R)=0$ and the NEC condition boils down to

\begin{equation} 
\left\{
\begin{array}{ll}
T_{kk} &= 0 \geq 0 \ , \\
T_{ll} &= -\frac{1}{32 \pi R} \frac{\partial\,F}{\partial v} \geq 0 \ .
\end{array}
\right.
\end{equation}
The NEC is thus satisfied if and only if $\partial_v F \leq 0$. Looking at the form of $F$ \eqref{metric_Hayward} and taking $R \rightarrow +\infty$, one recovers the standard ingoing Vaidya metric $F=1-\frac{2M(v)}{R}$. This means that the NEC is violated at infinity as soon as $M$ becomes a decreasing function of $v$. 
Using the previous calculations for various models, one sees in fact that there exists a line dividing the whole spacetime into a NEC-satisfying and a NEC-violating region (Fig.~\ref{fig_NEC_violation_Bardeen}, the NEC line represents $\partial_vF=0$). 

\subsection{Explicit EMT for the formation and evaporation of a non-singular trapped region}

\subsubsection{Conditions on the EMT}

The requirements on the EMT for obtaining a transition from the collapse of a null ingoing Vaidya shell to a Hayward-like non-singular trapped region, which then evaporates forming a null outgoing Vaidya shell on $\mathscr{I}^+$, are the following

\begin{equation}
\begin{aligned}
 T_{ll} \gg T_{kk},T_{kl}, T_{\theta \theta}  \text{ on } \mathscr{I}^- \ , \\
 T_{kk}  \gg T_{ll},T_{kl}, T_{\theta \theta}  \text{ on } \mathscr{I}^+ \ .
\end{aligned}
\end{equation}
We also demand that the null energy condition be satisfied up to infinity, thus:

\begin{equation}
T_{ll}, T_{kk}  \geq 0  \text{ on } \mathscr{I}^- \text{ and } \mathscr{I}^+ \ .
\end{equation}
The $\mathscr{I}^+$ and $\mathscr{I}^-$ limits are characterized by $v \rightarrow +\infty$ and $u \rightarrow -\infty$, where $u=v-2R$. We can define all functions in terms of $u$ and $v$, which gives

\begin{equation} \label{EMT_null_comp_uv}
\left\{
\begin{array}{lllll}
T_{kk}&=-\frac{e^{-2 \, \psi(u,v)} \partial_u \psi}{2\pi (v-u)} \ , \\
T_{ll}&=-\frac{F\left(u, v\right)^{2} e^{2 \, \psi\left(u, v \right)} \partial_u \psi + \, e^{\psi\left(u, v \right)} \frac{\partial\,F}{\partial v}}{32\pi (v-u)} \ , \\
T_{kl}&=\frac{ F\left(u, v\right)\partial_u \psi  + \partial_u F+ \frac{F(u,v)-1}{v-u}}{8 \pi (v-u)} \ .
\end{array}
\right.
\end{equation}
In order to obtain an explicit energy-momentum tensor describing the formation and evaporation of a non-singular trapped region, we will have the freedom to choose $\psi$. Indeed, this function will not affect the form of the horizons

\begin{equation}\label{psi_horizons}
\theta_+=\frac{F(v,R) \text{e}^{\psi(v,R)}}{R}=0 \Leftrightarrow F=0 \ .
\end{equation}
We can thus look for a function $\psi$ to model the gravitational collapse and Hawking radiation while keeping the horizons of the Hayward, Frolov or Bardeen cases; this is the purpose of the next two subsections.

\subsubsection{Choice of $\psi$ on $\mathscr{I}^+$}

Let us start by describing the phase of evaporation of the trapped region, mimicking the Hawking radiation by the energy-momentum tensor of an outgoing Vaidya metric. The component which must dominate all others is

\begin{equation}
T_{kk}=-\frac{e^{-2 \, \psi(u,v)} \partial_u \psi}{\pi (v-u)}. 
\end{equation}
It is thus clear that $\psi$ must not be a constant in order to obtain a flux of Hawking radiation on $\mathscr{I}^+$. For simplicity we can choose $\psi$ of the form $\psi=\psi(u)$. This allows avoiding a violation of the NEC on $\mathscr{I}^+$, as well as recovering Minkowski's metric there (up to a rescaling of the advanced time $v$). Furthermore, the intensity of the Hawking flux being driven by $T_{kk}$, hence by $\partial_u \psi$, we are looking for a function $\psi$ with an important slope for a given interval of $u$ (the phase of Hawking radiation) and which tends towards a constant value for large $u$. The following function meets all the above criteria:  

\begin{equation}
\psi(u)=\arctan \left(1000-u \right) \ .
\end{equation}
This leads to 

\begin{equation}
\begin{aligned}
T_{kk}&=\frac{ \exp (-2 \arctan (1000-u) )}{\pi (1 + (v- u)^2) (v-u)} \ , \\
T_{ll}&=-\frac{\exp(\arctan(1000-u))}{32\pi (v-u)} \left[\partial_v F-\frac{\exp( \arctan(1000-u))F^2}{1+(1000-u)^2} \right] \ .
\end{aligned}
\end{equation}
On $\mathscr{I}^+$, $v \rightarrow +\infty$ and we immediately have $T_{kk} \rightarrow 0^+$. As concerns $T_{ll}$, $\partial_v F \rightarrow 0$ and the second term thus dominates in the bracket. Hence, $T_{ll} \rightarrow 0^+$ on $\mathscr{I}^+$ as well. \\

\begin{figure}[h]
\centering
\begin{subfigure}{.6\textwidth}
  \centering
  \hspace{-2cm}
  \includegraphics[scale=0.55]{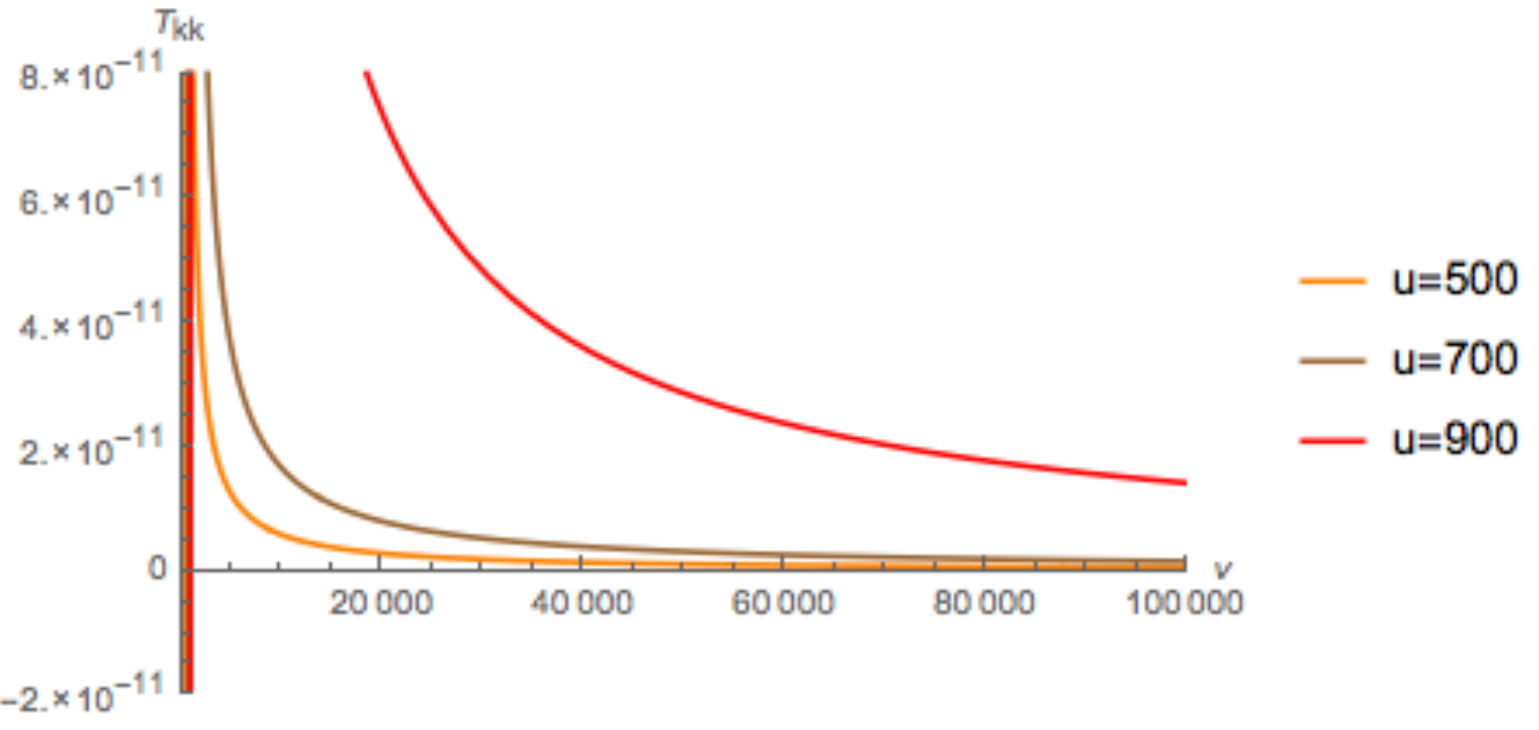}
  \label{fig:sub1}
\end{subfigure}%
\begin{subfigure}{.5\textwidth}
  \centering
    \hspace{-1cm}
    \includegraphics[scale=0.55]{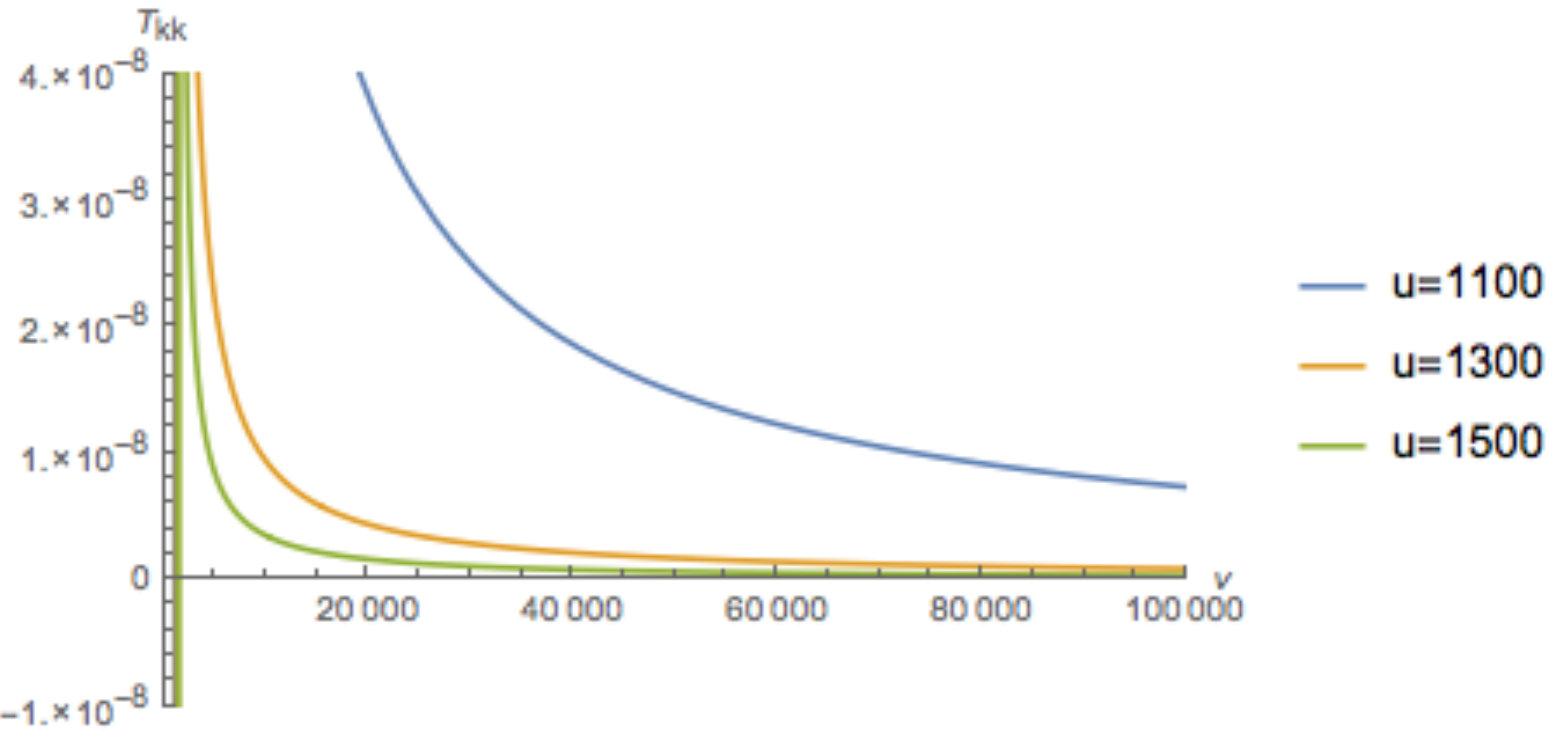}
  \label{fig:sub2}
\end{subfigure}
\caption{\footnotesize Plots of $T_{kk}$ as a function of $v$ for $u=500, 700, 900, 1100, 1300, 1500$.}
\label{fig:Tkk}
\end{figure}

On Fig.~\ref{fig:Tkk}, one can see that for large positive values of $v$, the biggest values of $T_{kk}$ are centered around $u=1000$.  

\begin{figure}[h]
\centering
\begin{subfigure}{.6\textwidth}
  \centering
  \hspace{-2cm}
  \includegraphics[scale=0.55]{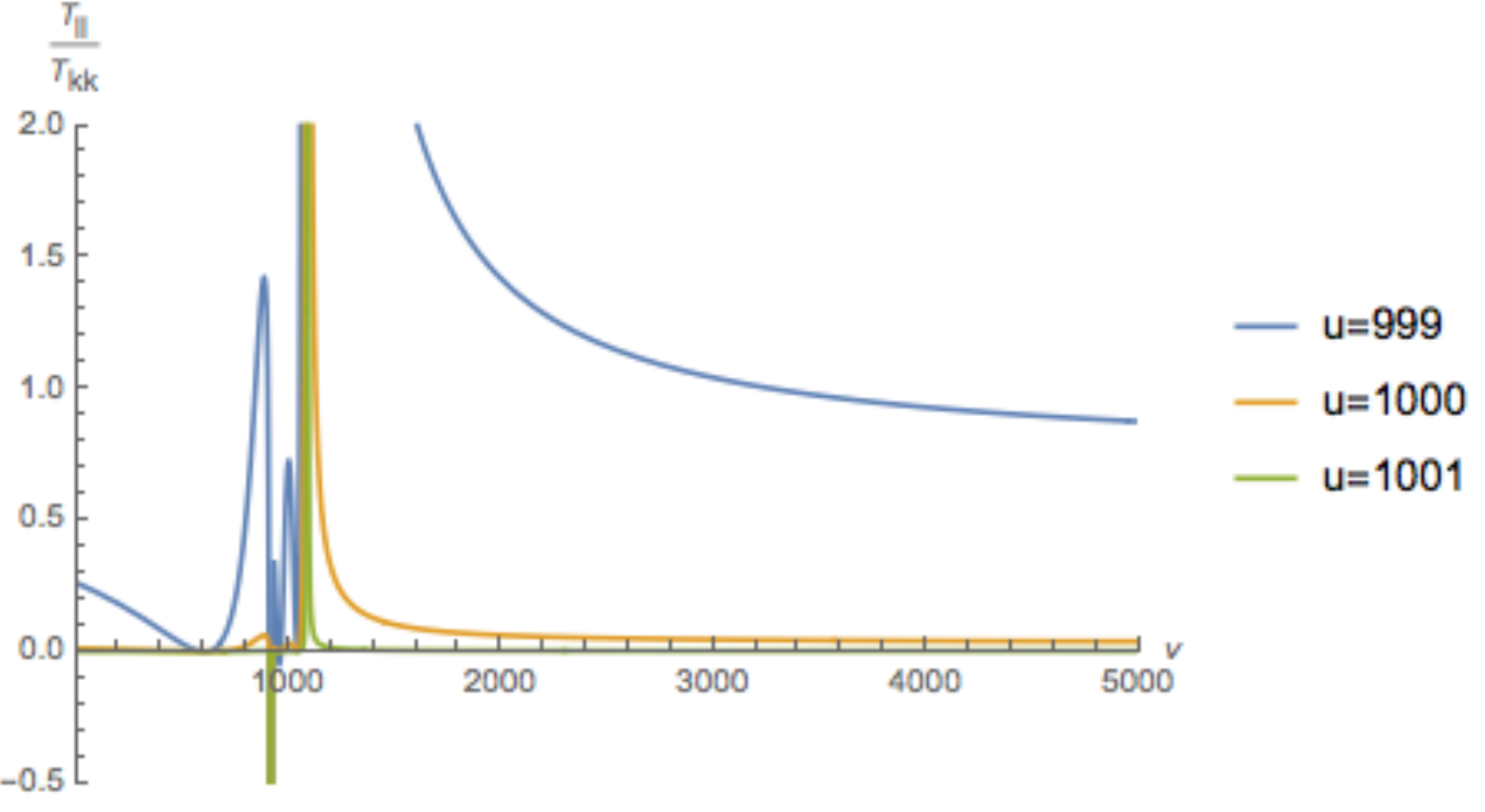}
  \label{fig:sub1}
\end{subfigure}%
\begin{subfigure}{.5\textwidth}
  \centering
    \hspace{-1cm}
    \includegraphics[scale=0.55]{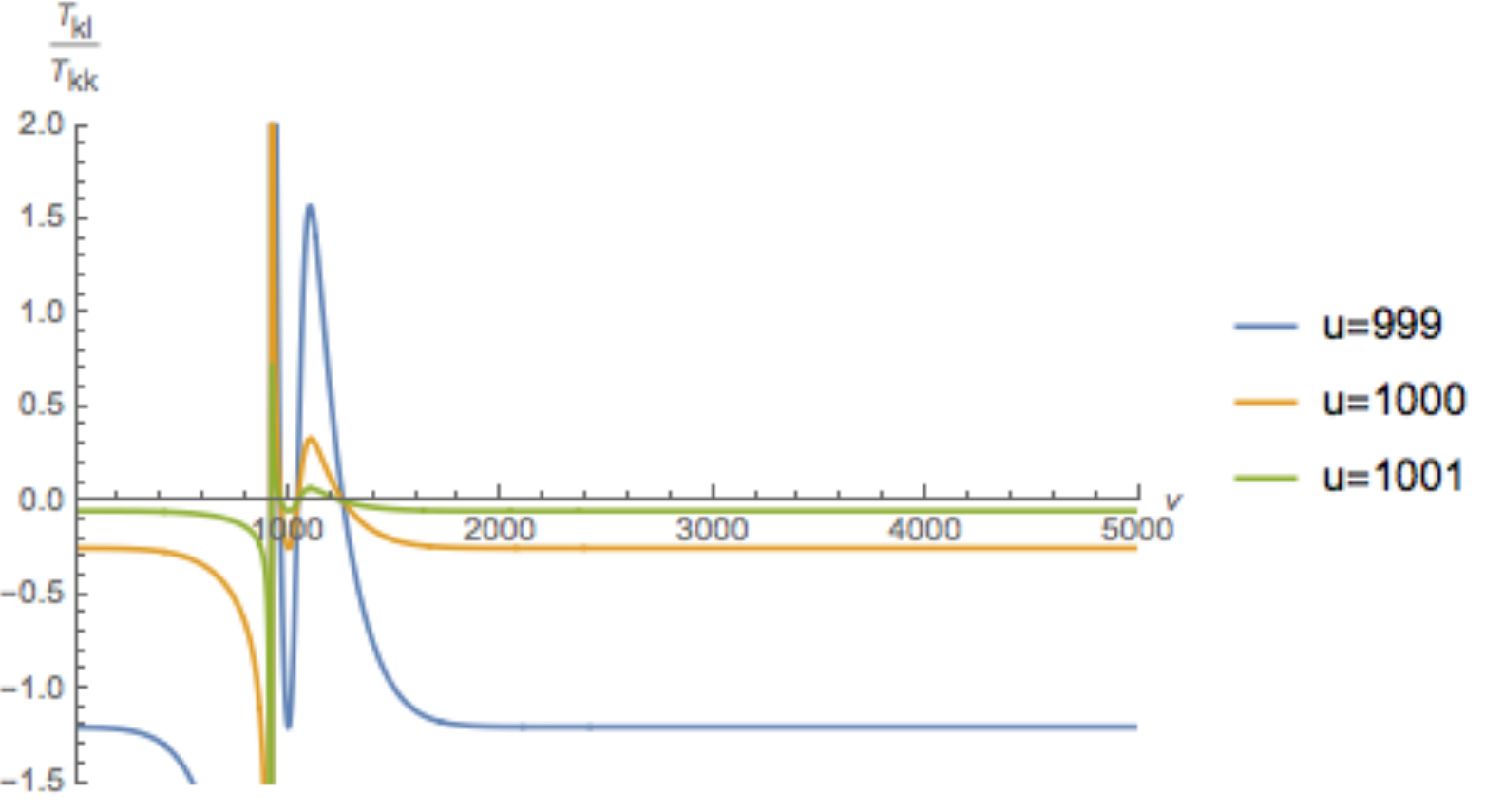}
  \label{fig:sub2}
\end{subfigure}
\caption{\footnotesize Plots of $\frac{T_{ll}}{T_{kk}}$ (left) and $\frac{T_{kl}}{T_{kk}}$ (right) as a function of $v$ for $u=999, 1000, 1001$.}
\label{fig:Tkk_dominates}
\end{figure}

Finally, it can be seen on Fig.~\ref{fig:Tkk_dominates} that abruptly after $u=1000$ and at large $v$, $T_{kk}$ dominates the other components of the energy-momentum tensor. This outgoing Vaidya-like behaviour mimicks the beginning of Hawking's radiation.

\subsubsection{Choice of $\psi$ on $\mathscr{I}^-$}

As explained in Section \ref{NEC violation}, choosing $\psi=0$ leads to a violation of the NEC on all $v=\text{cst}$ slices as soon as $m$ begins to decrease ($v>1000$ in the Hayward-like model). In order to avoid this violation of the NEC on $\mathscr{I}^-$, one has to carefully study the sign of the following components of the EMT:

\begin{equation}
\left\{
\begin{array}{lll}
T_{kk}&=-\frac{e^{-2 \, \psi(u,v)} \partial_u \psi}{\pi (v-u)} \ , \\
T_{ll}&=-\frac{e^{\psi\left(u, v \right)}}{32\pi (v-u)} \left[F^2 e^{\psi\left(u, v \right)}\partial_u \psi +\partial_v F \right] \ .
\end{array}
\right.
\end{equation}
First of all, choosing a function $\psi$ such that $\partial_u \psi \leq 0$ for $u \rightarrow -\infty$ ensures that $T_{kk}$ is non negative on $\mathscr{I}^-$. As concerns $T_{kk}$, the term inside the brackets must be negative for $u \rightarrow -\infty$. Let us study the sign of $\partial_v F$, assuming a form for $F$ boiling down to an ingoing Vaidya metric near  $\mathscr{I}^-$:

\begin{equation}\label{Vaidya_ing}
\left\{
\begin{array}{lll}
&F(v,R)=1-\frac{2M(v,R)}{R} \ , \\
 & \displaystyle \lim_{\substack{v=\text{cst}\\R \to +\infty}} M(v,R)=m(v) \ .
\end{array}
\right.
\end{equation}
Since $R=(v-u)/2$, the leading term of $\partial_v F$ on $\mathscr{I}^-$ is

\begin{equation}
\partial_v F \simeq \frac{4 \partial_v m(v)}{u} \ .
\end{equation}
Therefore, as long as $m(v)$ is increasing, $T_{ll} \rightarrow 0^+$ on $\mathscr{I}^-$ assuming that we keep on with $\psi$ such that $\partial_u \psi \leq 0$. \\

However, as soon as $m(v)$ becomes decreasing, this leads to $\partial_vF \rightarrow 0^+$ on $\mathscr{I}^-$. Since $\partial_u \psi \leq 0$, we have to study carefully the sign of $T_{ll}$. On $\mathscr{I}^-$, $F \rightarrow 1$ and we require that $e^{\psi\left(u, v \right)} \rightarrow \text{cst}$ so that we recover Minkowski's metric there. Hence, the study of the sign of $T_{ll}$ comes down to the comparison of the dominant terms of $\partial_u \psi$ and $\partial_v F \simeq \frac{4 \partial_v m(v)}{u}$. Ultimately, finding $\psi(u,v)$ boils down to satisfying simultaneously the three following	 conditions:

\begin{equation} 
\begin{aligned}
& \text{i)}\displaystyle \lim_{u \to -\infty} \psi(u,v)=a, a \in \mathbb{R} \ . \\
& \text{ii)} \lim_{u \to -\infty} \partial_u\psi(u,v)=b, b \in \mathbb{R}^- \ . \\
& \text{iii)} \text{ a) } \frac{1}{u} \underset{u \rightarrow -\infty}{=} o(\partial_u \psi(u,v)) \text{  or  } \text{b) } \frac{1}{u} \underset{u \rightarrow -\infty}{=} O(\partial_u \psi(u,v)) \ .
\end{aligned}
\end{equation}
Starting for instance from iii)a) and ii), we get

\begin{equation}
u \partial_u \psi(u,v) \underset{u \rightarrow -\infty}{\longrightarrow} +\infty \ . 
\end{equation}
Hence, for all $p \in \mathbb{R}^-$ there always exists large enough negative values $u$ such that $u \partial_u \psi(u,v) ~>~p$. Thus

\begin{equation}
\partial_u \psi(u,v) < \frac{p}{u} \ ,
\end{equation}
and 

\begin{equation}
\psi(u,v) < p \log(-u) \ .
\end{equation}
Hence $\psi(u,v)$ can be made arbitrarily large in absolute value, in contradiction with i). A similar reasoning applies with condition iii)b). \\
 
Finally, we have shown that every spacetime equipped with a metric of the ingoing Vaidya form \eqref{Vaidya_ing} near $\mathscr{I}^-$ will violate the NEC in a non-compact region as soon as $m(v)$ decreases. This applies, in particular, to the models of Hayward, Frolov and Bardeen, which mimick the Hawking evaporation through a decreasing function $m(v)$.

\section*{Conclusion}
In gravitational collapse forming black holes, the first MOTS usually forms inside the contracting matter, and then evolves into an inner and an outer trapping horizons. Classically, a singularity forms when the inner horizon reaches the centre of the cloud, while the outer horizon asymptotes to the usual event horizon. In other words, the appearance of a singularity stops the evolution of the inner trapping horizon, while the outer one eventually becomes isolated and null when all neighbouring matter has fallen in. It is usually thought that it is this horizon that has to be monitored in order to understand the behaviour and fate of the black hole, especially through the quantum emission of Hawking radiation; but the outer horizon is certainly not the only place to look for significant quantum effects.

It is clear that the status of the singularity is not well-defined when quantum effects are taken into account: they may very well regularize the classical singularity. In this case, there is currently no way to tell whether the inner horizon would still be stopped, or what subsequent evolution it would have; it could even be the locus of instabilities \cite{Frolov&Zelnikov:2017}. The possibility is therefore completely open for the evolution of the inner horizon to be greatly affected by these quantum effects, for instance to experience a bounce around the Planck scale as in the Bardeen-like model and be rejected to larger radii, eventually reaching the outer trapping horizon. This would lead to the vanishing of all spherically symmetric trapped surfaces in the spacetime, and to a very different picture from the usual black hole paradigm and the related information loss paradox. There would only be an asymptotically flat spacetime, with contracting matter forming a trapped region which would last for some time before disappearing. This would also be the case with the Hayward-like model where the outer horizon shrinks due to Hawking radiation, before joining the inner horizon at microscopic values. These scenarios of closed trapping horizons cannot be dismissed and have to be examined. 

After exposing this idea, we derived the minimal conditions for obtaining a non-singular spacetime with closed trapping horizons. We obtained a minimal form of the metric corresponding to a generalization of Hayward's metric \cite{Hayward:2005gi}. We then reviewed some important existing models \cite{Hayward:2005gi,Frolov:2014jva,Bardeen:2014uaa} within this framework, listing their advantages and limitations. In particular, they all display the physical limitation of having a violation of the NEC in a non-compact region up to null infinity as soon as the mass starts decreasing. We then studied the behaviour of radial null geodesics in such spacetimes, trying to identify the most relevant hypersurfaces in the study and characterization of these models. Finally, we endeavoured to build a spacetime of the above kin but without the physical limitations previously listed, and analyse the physical content of the energy-momentum tensor by solving Einstein's equations in reverse. We derived the conditions for avoiding a violation of the NEC in a non-compact region, and the requirements on the EMT at past and future null infinities for having a physical fluid satisfying the NEC. We found that, while it is possible to have the desired behaviour on  $\mathscr{I}^+$ via the introduction of a suitable function $\psi$, it is not on $\mathscr{I}^-$. Therefore, it is not possible to construct a non-singular spacetime with closed trapping horizons, asymptotically originating from a Vaidya collapse, that violates the NEC in a compact region only. In particular, the numerous proposals based on a Hayward-like metric will all share this limitation. Therefore, one would have to look for a more complicated collapse than the one starting from $\mathscr{I}^-$  with an ingoing null fluid in order to achieve the goal of having closed trapping horizons with a localized violation of the NEC.
\vspace{1cm}

\section*{Acknowledgements}

 In the process of preparing this paper, we have benefited from  
discussions with a number of colleagues; in particular, we are very grateful to David Langlois and \'{E}ric Gourgoulhon. We also want to thank David Vannerom, who was part of the early stages of this project, and Valeri Frolov for his kind suggestions. F. L. also wishes to thank Nathalie Deruelle and Cristina Volpe for their support in the last stages of this project. 
The research leading to these results has received funding from the ERC Advanced Grant 339169 ``Self-Completion''. \\ \par

Unfortunately, Pierre Binétruy could not participate in the completion of this work, neither could he explicitly endorse the substantial modifications that we made subsequently (especially in Section 4). He pursued the project in spite of everything, and we hope that he would have liked this final version.

\end{document}